\documentclass[english,preprint,tightenlines,nofootinbib,byrevtex,superscriptaddress]{revtex4}

\usepackage{graphicx}
\usepackage[latin1]{inputenc}
\usepackage{amsmath}
\usepackage{amssymb}
\usepackage{babel}
\usepackage{bm}
\usepackage{nicefrac}
\def\bra#1{\mathinner{\langle{#1}|}}
\def\ket#1{\mathinner{|{#1}\rangle}}
\def\braket#1{\mathinner{\langle{#1}\rangle}}

{\catcode`\|=\active 
  \gdef\Braket#1{\begingroup
     \ifx\SavedDoubleVert\relax
       \let\SavedDoubleVert\|\let\|\BraDoubleVert
     \fi
     \mathcode`\|32768\let|\BraVert
     \left<{#1}\right>\endgroup}
}
\def\BraVert{\@ifnextchar|{\|\@gobble}
     {\egroup\,\mid@vertical\,\bgroup}}
\def\BraDoubleVert{\egroup\,\mid@dblvertical\,\bgroup}
\let\SavedDoubleVert\relax

{\catcode`\|=\active
  \gdef\set#1{\mathinner{\lbrace\,{\mathcode`\|32768\let|\midvert #1}\,\rbrace}}
  \gdef\Set#1{\left\{
     \ifx\SavedDoubleVert\relax \let\SavedDoubleVert\|\fi
     \:{\let\|\SetDoubleVert
     \mathcode`\|32768\let|\SetVert
     #1}\:\right\}}
}
\def\midvert{\egroup\mid\bgroup}
\def\SetVert{\@ifnextchar|{\|\@gobble}
    {\egroup\;\mid@vertical\;\bgroup}}
\def\SetDoubleVert{\egroup\;\mid@dblvertical\;\bgroup}

\begingroup
 \edef\@tempa{\meaning\middle}
 \edef\@tempb{\string\middle}
\expandafter \endgroup \ifx\@tempa\@tempb
 \def\mid@vertical{\middle|}
 \def\mid@dblvertical{\middle\SavedDoubleVert}
\else
 \def\mid@vertical{\mskip1mu\vrule\mskip1mu}
 \def\mid@dblvertical{\mskip1mu\vrule\mskip2.5mu\vrule\mskip1mu}
\fi

\makeatletter
\newcommand{\be}{\begin{equation}}
\newcommand{\ee}{\end{equation}} 

\makeatother

\allowdisplaybreaks

\begin{document}

\title{Atomic Electric Dipole Moments: The Schiff Theorem and Its Corrections}

\author{C.-P. Liu}
\affiliation{T-16, Theoretical Division, Los Alamos National Laboratory, Los Alamos,
NM 87545, USA}
\affiliation{Theory Group, Kernfysisch Versneller Instituut, University of Groningen, \\
Zernikelaan 25, 9747 AA Groningen, The Netherlands}

\author{M. J. Ramsey-Musolf}
\affiliation{Kellogg Radiation Laboratory, California Institute of Technology, \\
Pasadena, CA 91125, USA}
\affiliation{Department of Physics, University of Wisconsin-Madison, Madison,
WI 53706, USA}

\author{W. C. Haxton}
\affiliation{Institute for Nuclear Theory and Department of Physics, University of Washington, \\
Box 351550, Seattle, WA 98195-1550, USA}

\author{R. G. E. Timmermans}
\affiliation{Theory Group, Kernfysisch Versneller Instituut, University of Groningen, \\
Zernikelaan 25, 9747 AA Groningen, The Netherlands}

\author{A. E. L. Dieperink}
\affiliation{Theory Group, Kernfysisch Versneller Instituut, University of Groningen, \\
Zernikelaan 25, 9747 AA Groningen, The Netherlands}

\preprint{LA-UR-07-2262, Caltech MAP-311}

\begin{abstract}
Searches for the permanent electric dipole moments (EDMs) of diamagnetic
atoms provide powerful probes of CP-violating hadronic and semileptonic
interactions. The theoretical interpretation of such experiments,
however, requires careful implementation of a well-known theorem by
Schiff that implies a vanishing net EDM for an atom built entirely
from point-like, nonrelativistic constituents that interact only electrostatically.
Any experimental observation of a nonzero atomic EDM would result
from corrections to the point-like, nonrelativistic, electrostatic
assumption. We reformulate Schiff's theorem at the operator level and delineate
the electronic and nuclear operators whose atomic matrix elements
generate corrections to ``Schiff screening''. We obtain a form
for the operator responsible for the leading correction associated
with finite nuclear size -- the so-called ``Schiff moment'' operator
-- and observe that it differs from the corresponding operator used
in previous Schiff moment computations. We show that the more general
Schiff moment operator reduces to the previously employed operator
only under certain approximations that are not generally justified.
We also identify other corrections to Schiff screening that may not
be included properly in previous theoretical treatments. We discuss
practical considerations for obtaining a complete computation of corrections
to Schiff screening in atomic EDM calculations.
\end{abstract}

\maketitle

\section{Introduction\,\label{sec:intro}}

The search for CP-violation (CPV) in and beyond the Standard Model
(SM) is important to particle, nuclear, and atomic physics and their
intersections with cosmology. The CPV in the electroweak sector of
the SM, parametrized by the complex phase in the Cabibbo-Kobayashi-Maskawa
(CKM) matrix, adequately accounts for experimental observations of
CPV effects in neutral kaon and $B$ meson systems. The results of searches
for the permanent electric dipole moments (EDMs) of neutron ($d_{n}$)
and diamagnetic atoms ($d_{A}$) imply that CPV in the strong sector
of the SM, parametrized by the so-called ``$\theta$-term'' are
vanishingly small: $|\bar{\theta}|\lesssim10^{-10}$\,\cite{Baker:2006ts,Romalis:2000mg}.
The so-called strong CP-problem associated with the unnaturally small
value of $\bar{\theta}$ may find resolution in spontaneously broken
Peccei-Quinn symmetry if the corresponding Goldstone boson -- the
axion -- is found. If so, strong CPV may hold the key to one of the
outstanding puzzles in cosmology, as the axion is a viable candidate
for particle dark matter.

Nevertheless, another cosmological problem -- the origin of the non-vanishing
baryon asymmetry of the universe (BAU) -- would remain unsolved. As
noted by Sakharov nearly four decades ago\,\cite{Sakharov:1967dj},
if the initial conditions of the universe were matter-antimatter symmetric,
the particle physics of the evolving universe would have to contain
CPV interactions (along with baryon number violation and a departure
from thermal equilibrium) in order to produce a BAU. It has been subsequently
noted that the CPV contained in the SM is not sufficiently effective
to produce the observed BAU, so an explanation requires new sources
of CPV. If such interactions involve particles having masses $\lesssim$
a few TeV, then current searches for atomic, hadronic, and leptonic
EDMs could soon yield nonzero results. Existing limits already provide
stringent constraints on new electroweak-scale CPV, and the next generation
of experiments are poised to push the sensitivity to new CPV sources
by several orders of magnitude. If new electroweak-scale CPV in models
like supersymmetry is, indeed, responsible for the BAU, then one would
expect nonzero results in this new round of EDM searches (see, \textit{e.g.},
Ref.\,\cite{Cirigliano:2006dg} and references therein. For a recent review of EDMs and CPV, see Ref.\,\cite{Pospelov:2005pr}).

If a nonzero EDM is observed, then it will take a set of complementary
EDM experiments -- together with the results of future collider searches
and precision electroweak studies -- to identify the nature of the
CPV. In this respect, searches for EDMs of diamagnetic atoms play
a key role. Experimentally, new efforts to probe $d_{A}$ with Ra\,\cite{-Ra_ANL,-Ra_KVI},
Xe\,\cite{-Xe_Princeton}, and Rn\,\cite{-Rn_TRIUMF} are presently
underway. The sensitivity of these atomic EDMs to either $|\bar{\theta}|$
or CPV beyond that of the SM could exceed that obtained for the $^{199}$Hg
atom\,\cite{Romalis:2000mg} , which currently sets the standard
for $d_{A}$, by several orders of magnitude. The $^{199}$Hg standard
is also expected to improve by an order of magnitude, when current
experiments are completed\,\cite{-Hg_Seattle}. Theoretically the
interpretation of $d_{A}$ is challenging because of the interplay
of atomic, nuclear, and hadronic physics. Specifically, it has long
been recognized that an EDM residing on the nucleus of a neutral diamagnetic
atom would not be detectable, apart from small corrections: Schiff\,\cite{Schiff:1963}
showed that any neutral system of electrically charged, nonrelativistic,
point-like constituents interacting only electrostatically will have
no net EDM.\,%
\footnote{In fact, part of this shielding effect has already been pointed out
by Purcell and Ramsey\,\cite{Purcell:1950} much earlier, but it
is Schiff who first clearly demonstrates this effect in quantum mechanics.
} %

Schiff's result can be understood at the classical level. An atomic
EDM is probed by placing the atom in a combination of external magnetic
and electric fields. If the atom has a net EDM, then its interaction
with the external electric field $\bm E_{{\rm ext}}$ would lead to
a shift in the Larmor precession frequency that depends linearly on
$\bm E_{{\rm ext}}$. However, a neutral system in the presence of
an external electric field cannot accelerate. Thus, the atomic constituents
-- assumed here to be point-like -- must rearrange so as to screen
out the effect of $\bm E_{{\rm ext}}$ at the location of each of
the charged constituents. Consequently, if any of the charged constituents
(electrons or nucleus) possesses a nonzero EDM, changes in the interaction
energy due to $\bm E_{{\rm ext}}$ will be exactly cancelled by those
from the internal fields induced by atomic polarization. To the extent
that the interaction of the atom with $\bm E_{{\rm ext}}$ arises
solely from the EDMs of its point-like, nonrelativistic constituents,
there will be no shift in the Larmor frequency linear in $\bm E_{{\rm ext}}$
and, hence, no effective atomic EDM. Corrections to this exact screening
will arise if any of the constituents having an EDM (a) moves relativistically,
leading to a breakdown of the classical picture, (b) has a finite
size, leading to a breakdown of the point-like assumption, or (c)
has non-electrostatic interactions that become modified in the presence
of the atomic rearrangement needed to achieve electrostatic equilibrium.
In addition, CPV interactions between constituents can lead to an
atomic EDM that evades the Schiff screening.

Because of such corrections and the exquisite sensitivity of the measurements,
atomic EDM experiments place important constraints on the electron
EDM and on CPV interactions in the nucleus. The present limits on
the electron EDM $d_{e}$, for example, are derived from heavy paramagnetic
atoms, where the Schiff theorem is evaded by the relativistic motion
of the atomic electrons. In the case of diamagnetic atoms, nonzero
effects due to hadronic CPV arise because of corrections (b) and (c),
specifically the finite size of the nucleus that leads to imperfect
shielding and magnetic and other, higher-order multipole interactions
between the nucleus and atomic electrons. The finite-size correction
becomes much more important in heavier atoms: to the extent that the
atomic electrons penetrate the atomic nucleus, a residual CPV interaction
arises. This penetration produces an atomic EDM proportional to the
so-called ``Schiff moment'' $\langle\hat{\bm S}\rangle$ of the nucleus,
often expressed as an $r^{3}$-weighted integral over the difference
in the nuclear charge and EDM distributions. Additional contributions
to the atomic EDM are generated by hyperfine (non-electrostatic) interactions
between atomic electrons and CP-conserving higher-order nuclear moments,
such as the $M_{1}$ (magnetic dipole) and $C_{2}$ (charge quadrupole),
as well as higher-order CPV nuclear moments that couple to spatially
varying fields, such as the $M_{2}$ (magnetic quadrupole).

These general features of Schiff screening and their corrections were
identified in Schiff's original paper, and subsequently discussed
in other work by Sandars\,\cite{Sandars:1968b}, Feinberg\,\cite{Feinberg:1977},
and Sushkov, Flambaum, and Khriplovich\,\cite{Sushkov:1984} (for
later discussions, see, \textit{e.g.}, Refs.\,\cite{Lindroth:1989a,Khriplovich:1997,Engel:1999np,Sandars:2001nq,Flambaum:2001gq}).
The formulation of the corrections to Schiff screening in Refs.\,\cite{Feinberg:1977,Sushkov:1984}
concentrated on the Schiff moment effect, and approximate expressions
for $\hat{\bm S}$ were derived. The latter have been used in subsequent
theoretical treatments of the EDMs of $^{199}$Hg and other atoms,
from which limits on CPV parameters, such as $|\bar{\theta}|$, have
been derived (see, \textit{e.g.} Ref.\,\cite{deJesus:2005nb} and
references therein). While these treatments are correct at a qualitative
level, more precise derivations are possible, and likely important
given the prospect that nonzero atomic EDMs will soon be seen. In
what follows, we provide a systematic derivation of Schiff's theorem
and associated corrections in a spherical multipole formalism. When
comparisons are made to earlier work, several refinements are apparent: 

\begin{itemize}
\item [(i)] Our approach expresses the electron-nucleus interaction in
terms of spherical multipole operators, separating contributions associated
with electron penetration from those associated with a point-like
nucleus. In so doing, we obtain a form for the atomic operator $\hat{O}_{{\rm Schiff}}^{{\rm atomic}}$
that describes the leading corrections to Schiff screening associated
with electronic penetration into the nucleus. This atomic operator
can be applied to any atomic EDM and gives a more general characterization
of the leading finite size corrections than has been used in previous
studies.
\item [(ii)] We subsequently derive the nuclear Schiff moment operator
$\hat{\bm S}$ by considering matrix elements of $\hat{O}_{{\rm Schiff}}^{{\rm atomic}}$
between electronic S- and P-states as is relevant for many atomic
EDMs. Our result for $\hat{\bm S}$, the analog of the Schiff moment
operator used in many earlier studies, has a more general form than
the one in common use. We discuss the approximations under which our
expression for $\hat{\bm S}$ reduces to the form conventionally employed,
and observe that these approximations are not generally justified.
We advocate the use of the operator $\hat{\bm S}$ derived below that
does not require adoption of any approximations. We also note that
the operator $\hat{\bm S}$ derived for electronic S--P transitions
is not appropriate for other transitions, such as the close-lying
D-P transition in a metastable Ra atom where a large enhancement factor
is found\,\cite{Dzuba:2002kg}.
\item [(iii)] We derive expressions for the operators that characterize
interactions between the polarized atomic cloud and CP-conserving
nuclear moments, such as the magnetic dipole and charge quadrupole
moments. We also include finite-size, electronic penetration effects
associated with these multipole moments.
\item [(iv)] We obtain additional electronic penetration corrections involving
higher CPV multipole moments of the nuclear charge and current densities,
such as the charge octupole moment and magnetic quadrupole moment.
\item [(v)] We revisit corrections to Schiff screening of electron EDMs
and include corrections associated with both the external and nuclear
vector potentials that have not been included in earlier studies.
\end{itemize}
In addition to deriving the operators associated with the foregoing
effects, we also discuss practical considerations involved in computing
atomic EDMs. For example, a complete computation of effects associated
with hadronic CPV would require simultaneous diagonalization of the
electronic and nuclear Hamiltonian, taking into account effects such
as the nuclear response to the polarized atomic cloud and the corresponding
interaction with the external electric field. Practically speaking,
carrying out such diagonalization of the full space of atomic wave
functions is not possible even in the limit of purely CP-conserving
interactions. In typical atomic calculations, the nuclear charge and
current distributions are taken as $c$-number sources to which the
atomic electrons respond, and ``back reactions'' of the nuclear
charge and current distributions to the atomic electrons -- the so-called
nuclear polarization corrections -- are neglected. This procedure
amounts to treating the atomic states as direct products of the nuclear
ground state and different electronic states, with the properties
of the nuclear ground state acting as a $c$-number input for diagonalization
of the electronic Hamiltonian:
\begin{align}
\ket{n}_{\mathrm{atom}} & =\ket{\mathrm{g.s.}}_{\mathcal{N}}\otimes\ket{k}_{e}\,,
\label{eq:fact}
\end{align}
where the ${\cal N}$ and $e$ subscripts refer to the nuclear ground
state and electronic states (labeled by quantum numbers $k$), respectively.

One would na{\"i}vely expect the nuclear polarization corrections
to atomic properties associated with this factorization to be small,
going as powers of the nuclear radius, $R_{{\cal N}}$, divided by
the effective Bohr radius, $a_{0}$. For electronic S-states, however,
the situation is more subtle. Theoretical studies of muonic atoms,
for which the effective Bohr radius is $\sim200$ times smaller than
for ordinary atoms, indicate that corrections to the energy splitting
between the 2S$_{1/2}$ and 2P$_{3/2}$ states depends on the square
of the muon wave function at the origin and an integral over the nuclear
photoabsorption cross section\,\cite{Bernabeu:1973uf,Bernabeu:1975ka,Friar:1977cf}.
Numerically, the theoretical polarization correction to $\Delta E(2\textrm{P}_{3/2}-2\textrm{S}_{1/2})$
in $(\mu\,^{4}\textrm{He})^{+}$ is about 0.2\% of the measured splitting.
In the case of $(e\,^{4}\textrm{He})^{+}$, the authors of Refs.\,\cite{Bernabeu:1973uf,Bernabeu:1975ka}
find that the nuclear polarization correction to the S-P splitting
is a few $\times10^{-4}$ times smaller in magnitude than the corresponding
finite size correction, which depends on the square of the electronic
wave function at the origin times the mean square nuclear charge radius.

The Schiff moment interaction itself characterizes finite-size corrections
of order $(R_{{\cal N}}/a_{0})^{2}$ relative to the energy associated
with the interaction of the atomic electrons with the EDM of a point-like
nucleus. Based on the studies of Refs.\,\cite{Bernabeu:1973uf,Bernabeu:1975ka,Friar:1977cf},
one would expect the nuclear polarization corrections to the finite-size 
effects characterized by the Schiff moment to be negligible.
In the present analysis, however, we have not quantified these polarization
effects, and leave this task to future work. Instead, we will proceed
by formulating the Schiff theorem at the operator level as far as
possible without making explicit reference to the atomic states or
invoking the factorization approximation, thereby avoiding the issue
of nuclear polarization corrections. Only when attempting to compare our Schiff
moment operator with the corresponding operator used previously in
the literature will we adopt the factorization \textit{ansatz} of
Eq.\,(\ref{eq:fact}). Even in this case, we find substantial differences
with Schiff moment operator used in earlier analyses.

With the foregoing caveats in mind, we believe that our reformulation
of Schiff's theorem represents a useful refinement that will allow
for a more complete inclusion of hadronic and nuclear structure contributions
to $d_{A}$ and provide for a sharper confrontation between theory
and experiment. We organize our reformulation in the remainder of
the paper as follows. In Section\,\ref{sec:formulation} we give
our multipole operator formulation of the theorem, including a delineation
of the various operators that characterize corrections to Schiff screening.
For completeness, we include here a discussion of both the hadronic
CPV effects as well as those associated with electronic EDMs. We include
here for the first time the corrections associated with non-electrostatic
interactions between atomic electrons and both nuclear and external
sources. In Section\,\ref{sec:schiff}, we obtain the full atomic
Schiff moment operator $\hat{O}_{{\rm Schiff}}^{{\rm atomic}}$ that
is independent of the atomic states and use it, together with the
factorization \textit{ansatz}, to derive an effective nuclear Schiff
moment operator $\hat{\bm S}$ that can be compared with the operator
used previously in the literature. We also derive the operator that
characterizes the leading corrections to Schiff screening associated
with magnetic interactions between the electrons and nucleus. In Section\,\ref{sec:practical},
we summarize our results and discuss the practical implementation
of Schiff's theorem including the issues alluded above. A few technical
details are given in the Appendices.

\section{Schiff's Theorem: A multipole operator formulation\,\label{sec:formulation}}

The interaction of an external field with the atomic EDM is odd under
both parity (P) and time-reversal (T) transformations but even under
charge conjugation (C). Assuming CPT invariance, as we do throughout
this paper, the observation of a nonzero $d_{A}$-induced frequency
shift is, thus, equivalent to the observation of CP-violation. Henceforth,
we will refer to atomic, electronic, and nuclear moments that violate
both P and T symmetries as PVTV moments, while those that respect
these symmetries are PCTC moments. Effects that violate P but conserve
T, such as the nuclear anapole moment, are also of considerable interest,
but we do not consider these effects here (for recent reviews, see
Refs.\,\cite{Haxton:2001mi,Haxton:2001ay}). Effects that conserve
P but violate T, which in combination with the weak interaction can
induce PVTV interactions, are also not explicitly treated here.

\subsection{The atomic Hamiltonian\,\label{subsec:atomic}}

To evaluate the consequences of PVTV interaction of atomic properties,
we separate the complete atomic Hamiltonian into PCTC ($H_{0}$) and
PVTV ($H_{I}$) terms, treating the latter as a perturbation: 
\begin{equation}
H_{\textrm{atom}}=H_{0}+H_{I}\,.
\label{eq:hatom}
\end{equation}
 We also include the interaction of an external field with the atom
in $H_{I}$.\,%
\footnote{As we are interested on permanent EDMs instead of induced EDMs, it
is the first-order Stark effect that one is after. Therefore, the
weak external field limit is adequate.} %
Before identifying the specific operators in $H_0$ and $H_I$, it is useful to delineate the different interactions between atomic electrons, the electrons and nucleus, and external fields and the atomic constituents:
\begin{enumerate}
\item The interactions between atomic electrons is
\be
H_{\mathrm{int}}^{(ee)} =V_{\mathrm{int}}^{(ee)} + {\tilde V}_{\mathrm{int}}^{(\tilde{e}e)}\,,
\ee
where we adopt a notation (used throughout this paper) where the superscript $(ee)$
denotes a PCTC electron-electron interaction with PCTC couplings on both electrons, while $(\tilde{e}e)$
denotes a PVTV interaction that arises from a PCTC coupling on one electron
and a PVTV coupling on the other.  Specifically, $V_{\mathrm{int}}^{(ee)}$ denotes the PCTC Coulomb and Breit interactions between electrons
\be
\label{eq:pctcee}
V_{\mathrm{int}}^{(ee)} = \frac{\alpha}{2}\, \sum_{i=1}^{Z}\, \,(\phi_{i}^{(e)}-\bm\alpha_{i}\cdot\bm A_{i}^{(e)})
\ee
and ${\tilde V}_{\mathrm{int}}^{(\tilde{e}e)}$ denotes the PVTV electron-electron interaction, which we write as
\be
\label{eq:pvtvee}
\tilde{V}_{\mathrm{int}}^{(\tilde{e}e)}=  \frac{\alpha}{2}\,\sum_{i=1}^{Z}\, d_{e}\,\beta\,\left[\bm\sigma_{i}\cdot\bm E_{i}^{(e)}+i\,\bm\alpha_{i}\cdot
\bm B_{i}^{(e)}\right] + \cdots \,.
\ee
The term given explicitly is the interaction of each electron's EDM with the electric and magnetic fields ($\bm E_{i}^{(e)}$,$\bm B_{i}^{(e)}$) created by the other electrons.  In addition, the ``$+\cdots$'' indicates that possible additional terms could be added to this, new exchanges generating PVTV interactions.
Here, $\alpha$ is the fine structure constant and $\beta_{i}$ and $\bm\alpha_{i}$
are the Dirac matrices acting on the $i$th electron.  The scalar potential 
$\phi_{i}^{(e)} = \phi^{(e)}(\bm x_{i})$ is the potential exerted by the other $Z-1$ electrons at the position $\bm x_{i}$
of the $i$th electron; the notation is similar for the vector potential $\bm A_{i}^{(e)}$, and for
the resulting electric and magnetic fields $\bm E_{i}^{(e)}$ and $\bm B_{i}^{(e)}$. The
$1/2$ factor in the electron-electron interaction is introduced to
ensure that one sums over distinct pairwise interactions.\,%
\footnote{In what follows, we factor out electric charge from the scalar and
vector potentials.} %
\item The interaction between the electrons and nucleus ($\mathcal{N}$) is
\be
H_{\mathrm{int}}^{(e\mathcal{N})} =-\alpha\, \sum_{i=1}^Z\, \left[\phi_i^{(\mathcal{N})}-
\bm\alpha_{i}\cdot\bm A_{i}^{(\mathcal{N})}\right] + \tilde{V}_{\mathrm{int}}^{(\tilde{e}\mathcal{N})}\,,
\ee
where $\phi_i^{(\mathcal{N})}$ and $\bm A_{i}^{(\mathcal{N})}$ are the scalar and vector potentials of the nucleus at the position of the $i$th electron.  These potentials, as discussed below, in
general contain both PCTC and PVTV contributions.  The second term,
$\tilde{V}_{\mathrm{int}}^{(\tilde{e}\mathcal{N})}$, is the PVTV contribution due to interaction of electron EDMs with the PCTC electric and magnetic fields ($\bm E_{i}^{(\mathcal{N})}$,$\bm B_{i}^{(\mathcal{N})}$) associated with the nuclear potentials,
\be
\tilde{V}_{\mathrm{int}}^{(\tilde{e}\mathcal{N})} = -\alpha\,\sum_{i=1}^{Z}\, d_{e}\,\beta\,\left[\bm\sigma_{i}\cdot\bm E_{i}^{(\mathcal{N})}+i\,\bm\alpha_{i}\cdot
\bm B_{i}^{(\mathcal{N})}\right] + \cdots \,.
\label{eq:vinteNt}
\ee
Again, the ``$+ \cdots$'' indicates additional PVTV non-electromagnetic
exchanges that might exist between the electrons and nucleus.

\item The interaction of the electrons with the applied (``external'') potentials $(\phi_{i}^{(ext)},{\bm A}_{i}^{(ext)})$ and corresponding fields ($\bm E_{i}^{({\mathrm ext})}$,$\bm B_{i}^{({\mathrm ext})}$)
\be
H_{\mathrm{ext}}^{(e)} = V_{\mathrm{ext}}^{(e)} + {\tilde V}_{\mathrm{ext}}^{(\tilde{e})}\,,
\ee
where 
\begin{align}
V_{\mathrm{ext}}^{(e)} & =-\alpha\,\sum_{i=1}^{Z}\,\left(\phi_{i}^{(ext)}-\bm\alpha_{i}\cdot\bm A_{i}^{(ext)}\right)\,,\\
\tilde{V}_{\mathrm{ext}}^{(\tilde{e})} & =  -\alpha\,\sum_{i=1}^{Z}\, d_{e}\,\beta\,\left(\bm\sigma_{i}\cdot\bm E_{i}^{(ext)}+i\,\bm\alpha_{i}\cdot\bm B_{i}^{(ext)}\right)\,.
\end{align}
\item The interaction of the nucleus with the external potentials
\be
\label{eq:hextnuc}
H_{\mathrm{ext}}^{(\mathcal{N})} = \alpha \int\, d^3y\, \left[\hat\rho^{(\mathcal{N})}(\bm y)\,\phi^{(ext)}(\bm y)- \hat{\bm j}^{(\mathcal{N})}(\bm y)\cdot{\bm A}^{(ext)}(\bm y)\right]\,,
\ee
where the hat over the nuclear potentials indicates that they are operators rather than $c$-number functions.  Again, the nuclear charge and three-current operators include a variety of terms
that can contribute to PCTC and PVTV interactions.  A multipole
expansion is helpful in separating the PCTC and PVTV terms.
\end{enumerate}

\noindent{\em The spherical multipole expansion}
\vskip 0.1in

To divide $H_{\mathrm{int}}^{(e\mathcal{N})}$ and $H_{\mathrm ext}^{(\mathcal{N})}$
into PCTC and PVTV interactions, we decompose the
potentials,
\begin{eqnarray}
\label{eq:phinuc}
\phi^{(\mathcal{N})}(\bm x) & = & \int\, d^{3}y\,\frac{\hat{\rho}^{(\mathcal{N})}(\bm y)}{|\bm x-\bm y|}\,,\\
\label{eq:anuc}
\bm\alpha\cdot\bm A^{(\mathcal{N})}(\bm x) & = & \int\, d^{3}y\,\frac{\bm\alpha\cdot\hat{\bm j}^{(\mathcal{N})}(\bm y)}{|\bm x-\bm y|}\,,
\end{eqnarray}
into multipoles.  One expands the photon Green's function in terms of spherical harmonics $Y_{J}^{M}$,
\begin{equation}
\frac{1}{|\bm x-\bm y|}=\sum_{J\ge0}\,\frac{4\,\pi}{2\, J+1}\,\left[\theta(x-y)\frac{y^{J}}{x^{J+1}}+\theta(y-x)\frac{x^{J}}{y^{J+1}}\right]\, Y_{J}(\hat{x})\odot Y_{J}(\hat{y})\,,
\label{eq:greenexp}
\end{equation}
where
\be
Y_{J}(\hat{x})\odot Y_{J}(\hat{y})  \equiv\sum_{M}\, Y_{J}^{M*}(\hat{x})\, Y_{J}^{M}(\hat{y})\,.
\ee
The resulting expressions for the potentials are
\begin{eqnarray}
\phi^{(\mathcal{N})}(\bm x) & = & \int\, d^{3}y\,\frac{\hat{\rho}^{(\mathcal{N})}(\bm y)}{|\bm x-\bm y|}\nonumber \\
& = & \sum_{J\ge0}\,\frac{4\,\pi}{2\, J+1}\,\frac{1}{x^{J+1}}\, Y_{J}(\hat{x})\odot[\hat{C}_{J}+\hat{\mathcal{C}}_{J}(x)]\,,
\label{eq:phimult}\\
\bm\alpha\cdot\bm A^{(\mathcal{N})}(\bm x)& = & \int\, d^{3}y\,\frac{\bm\alpha\cdot\hat{\bm j}^{(\mathcal{N})}(\bm y)}{|\bm x-\bm y|}\nonumber \\
& = & -\sum_{J\ge1}\,\frac{4\,\pi}{2\, J+1}\,\frac{1}{x^{J+1}}\,[Y_{J}(\hat{x})\otimes\bm\alpha]_{J}\odot[\hat{M}_{J}+\hat{\mathcal{M}}_{J}(x)]+\cdots \,.
\label{eq:amult}
\end{eqnarray}
 %
Here $[A\otimes B]_{J}^{M}$
is the standard notation for the coupling two spherical tensors $A$
and $B$ to a tensor of rank $J$; and the ``$+\cdots$'' indicate
the contributions from the transverse electric multipoles that will not contribute
to either PCTC or PVTV moments due to their P and T transformation properties.\,%
\footnote{For example, the $J=1$ transverse electric multipole moment gives
the nuclear anapole moment \,\cite{Haxton:1989ap}.} %
The nuclear vector potential $\bm A^{(\mathcal{N})}$ satisfies
the Coulomb gauge.

The charge ($\hat{C}$ and $\hat{\mathcal{C}}$) and transverse magnetic
($\hat{M}$ and $\hat{\mathcal{M}}$) nuclear multipole operators are defined
as
\begin{subequations}
\begin{eqnarray}
\hat{C}_{J}^{M} & = & \int\, d^{3}y\, y^{J}\, Y_{J}^{M}(\hat{y})\,\hat{\rho}(\bm y)\,,\\
\hat{M}_{J}^{M} & = & \int\, d^{3}y\,\bm[y^{J}\, Y_{J}(\hat{y})\otimes\hat{\bm j}(\bm y)]_{J}^{M}\,,\\
\hat{\mathcal{C}}_{J}^{M}(x) & = & \int\, d^{3}y\,\theta(y-x)\left[(x/y)^{2\, J+1}-1\right]\, y^{J}\, Y_{J}^{M}(\hat{y})\,\hat{\rho}(\bm y)\equiv\hat{\mathcal{C}}_{J<}^{M}(x)-\hat{\mathcal{C}}_{J>}^{M}(x)\,,\\
\nonumber
\hat{\mathcal{M}}_{J}^{M}(x) & = & \int\, d^{3}y\,\theta(y-x)\left[(x/y)^{2\, J+1}-1\right]\,\bm[y^{J}\, Y_{J}(\hat{y})\otimes\hat{\bm j}(\bm y)]_{J}^{M}\\
&& \equiv\hat{\mathcal{M}}_{J<}^{M}(x)-\hat{\mathcal{M}}_{J>}^{M}(x)\,.
\label{eq:multdef}
\end{eqnarray}
\end{subequations}
In the definitions of $\hat{\mathcal{C}}(x)$
and $\hat{\mathcal{M}}(x)$ -- the penetration terms that account for contributions when
the electron cloud is inside the nucleus -- the subscripts {}``$<"$ and {}``$>$''
refer to the parts containing $(x/y)^{2\, J+1}$ and $1$, respectively.

The moments (diagonal matrix elements) of multipole operators in Eqs.\.(\ref{eq:phimult}--\ref{eq:multdef}) have definite parity and time-reversal characters depending on their angular momentum, as indicated in Table\,\ref{tab:mult}. For completeness, we have also included the moments of the transverse electric multipole operators, $\hat{E}_J^M$ and $\hat{\mathcal{E}}_J^M(x)$. These are not relevant in PVTV calculations in which only nuclear moments are considered.
A PVTV nuclear moment can arise if and only if some source of parity and time-reversal violation is introduced. Thus nonzero moments corresponding to odd charge multipoles or even magnetic multipoles are the signatures of PVTV interactions.

\begin{table}
\begin{ruledtabular}
\begin{tabular}{ccccc}
Multipole Moment & PCTC & PVTV & PVTC & PCTV \\
&&&&\\
$C_J^M$,\,  $\mathcal{C}_J^M(x)$ & even $J$ & odd $J$ & x & x \\
&&&&\\
$M_J^M$,\,  $\mathcal{M}_J^M(x)$ & odd $J$ & even $J$ & x & x \\
&&&&\\
$E_J^M$,\,  $\mathcal{E}_J^M(x)$ & x & x & odd $J$ & even $J$
\end{tabular}
\end{ruledtabular}
\caption{Parity and time-reversal characters of the Coulomb ($C_J^M$), magnetic
($M_J^M$), and transverse electric ($E_J^M$) multipole moments.}
\label{tab:mult}
\end{table}

The electron-nucleus interaction can be divided formally into PCTC and PVTV components:
\be
\label{eq:enint}
H_{\mathrm{int}}^{(e\mathcal{N})} = -\alpha\, \sum_{i=1}^Z\, \left[\phi_i^{(\mathcal{N})}-
\bm\alpha_{i}\cdot\bm A_{i}^{(\mathcal{N})}\right]
+ \tilde{V}_{\mathrm{int}}^{(\tilde{e}\mathcal{N})} \equiv V_{\mathrm{int}}^{(e\mathcal{N})}+\tilde{V}_{\mathrm{int}}^{(e\tilde{\mathcal{N}})}+
\tilde{V}_{\mathrm{int}}^{(\tilde{e}\mathcal{N})}\,,
\ee
where $\tilde{V}_{\mathrm{int}}^{(\tilde{e}\mathcal{N})}$ was defined in Eq.\,(\ref{eq:vinteNt}).  In general the multipole decomposition is helpful in separating the first
two terms.  For example, in the case most often of interest in which a nuclear moment is
being taken
\begin{eqnarray}
V_{\mathrm{int}}^{(e\mathcal{N})} & \stackrel{\mathrm{diagonal}}{\longrightarrow} & -\alpha\, \sum_{i=1}^Z\, \left[ \sum_{J\ge0,\mathrm{even}}\,\frac{4\,\pi}{2\, J+1}\,\frac{1}{x_i^{J+1}}\, Y_{J}(\hat{x}_i)\odot[\hat{C}_{J}+\hat{\mathcal{C}}_{J}(x_i)] \right. \nonumber \\
& & + \left. \sum_{J\ge1,\mathrm{odd}}\,\frac{4\,\pi}{2\,J+1}\,\frac{1}{x_i^{J+1}}\,[Y_{J}(\hat{x}_i)\otimes\bm\alpha]_{J}\odot[\hat{M}_{J}+\hat{\mathcal{M}}_{J}(x_i)] \right]
\end{eqnarray}
\begin{eqnarray}
V_{\mathrm{int}}^{(e\tilde{\mathcal{N}})} & \stackrel{\mathrm{diagonal}}{\longrightarrow} & -\alpha\, \sum_{i=1}^Z\, \left[ \sum_{J\ge1,\mathrm{odd}}\,\frac{4\,\pi}{2\, J+1}\,\frac{1}{x_i^{J+1}}\, Y_{J}(\hat{x}_i)\odot[\hat{C}_{J}+\hat{\mathcal{C}}_{J}(x_i)] \right. \nonumber \\
& & + \left. \sum_{J\ge2,\mathrm{even}}\,\frac{4\,\pi}{2\,J+1}\,\frac{1}{x_i^{J+1}}\,[Y_{J}(\hat{x}_i)\otimes\bm\alpha]_{J}\odot[\hat{M}_{J}+\hat{\mathcal{M}}_{J}(x_i)] \right].
\label{eq:eNnew}
\end{eqnarray}
Thus one can identify the total electron-nucleus PVTV interaction $\tilde{V}_{\mathrm{int}}^{(e\tilde{\mathcal{N}}+\tilde{e}\mathcal{N})}$, the sum of terms where the PVTV coupling is either on the nucleus or on the electrons
\be
\tilde{V}_{\mathrm{int}}^{(e\tilde{\mathcal{N}}+\tilde{e}\mathcal{N})}=\tilde{V}_{\mathrm{int}}^{(e\tilde{\mathcal{N}})}
+\tilde{V}_{\mathrm{int}}^{(\tilde{e}\mathcal{N})}\,.
\ee

\begin{figure*}
\caption{The contributions to a nuclear EDM from (a) nucleon EDMs, (b) PVTV exchange currents, and (c) parity admixtures induced by PVTV $NN$ interactions.}
\label{fig:dNuc}

\begin{center}
\begin{tabular}{cccc}
\includegraphics[scale=0.65]{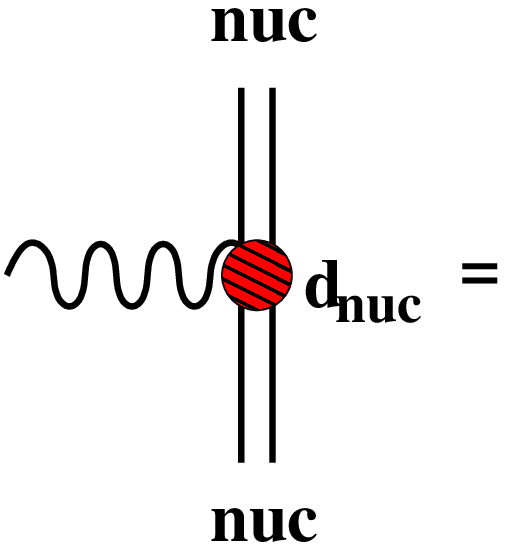}&
\includegraphics[scale=0.65]{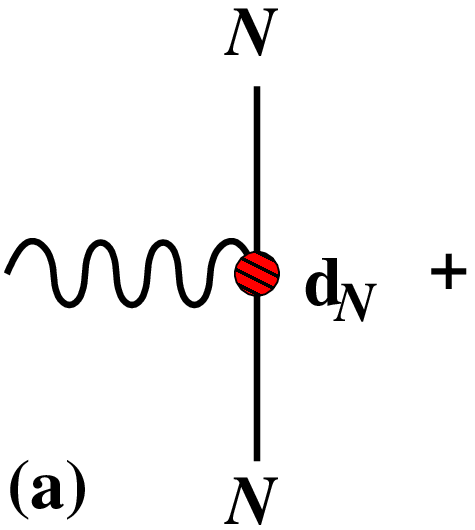}&
\includegraphics[scale=0.65]{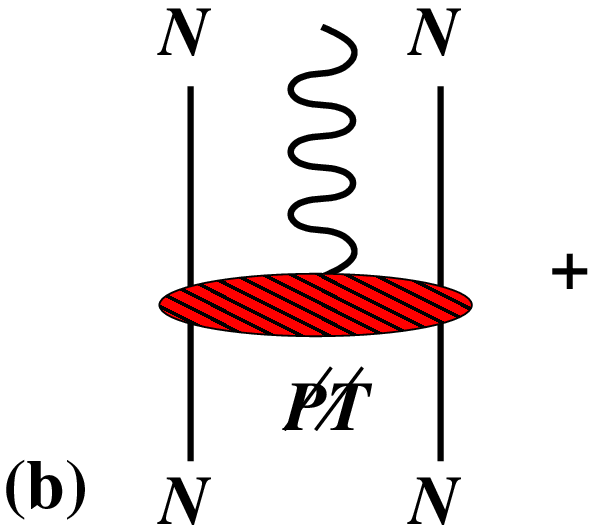}&
\includegraphics[scale=0.65]{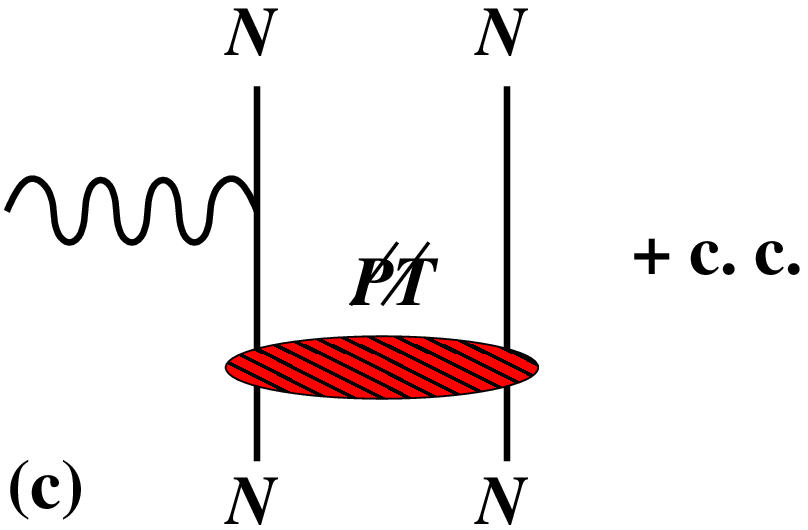}
\end{tabular}
\end{center}
\end{figure*}

In connection with Eq.\,(\ref{eq:eNnew}), it might be helpful at this point to comment about nuclear sources of PVTV interactions, which can originate from direct PVTV electromagnetic couplings to the nucleon (that is,
one-body PVTV currents generated by nucleon EDMs), couplings to various meson exchanges or to $N \bar{N}$ excitations (two-body PVTV currents), or PVTV interactions of any type between the
nucleons.  The latter would give rise to small PVTV admixtures in nuclear wave functions.
Thus we see, working to first order in PVTV, that there are two classes of contributions to PVTV
nuclear moments (see, {\it e.g.}, Refs.\,\cite{Herczeg:1966,Liu:2004tq} and references
therein):
\begin{itemize}
\item The PVTV current contribution: small components of the multipole operators due to PVTV contributions to one- or two-body charges or three-currents -- operators that might be denoted
$\hat{\tilde{C}}_{J=\mathrm{odd}}$ and $\hat{\tilde{M}}_{J=\mathrm{even}}$ --
would have nonzero matrix elements between the dominant PCTC component of the nuclear
wave function.
\item The PVTV nuclear polarization contribution: large components of the multipole operators due to the ordinary PCTC currents would have nonzero matrix elements connecting PCTC components in the bra state with PVTV ``polarization" admixtures in the ket state, and vice versa.
\end{itemize}
These contributions to the nuclear EDM are illustrated in Fig.\,\ref{fig:dNuc}.

In the multipole expressions (Eqs.\,(\ref{eq:phimult}-\ref{eq:eNnew})), we have been careful to
isolate terms associated with the finite nuclear size (and thus electron penetration) from
those that would persist in the point-nucleus limit.
Nuclear matrix elements of the operators $\hat{C}_{J}$'s and $\hat{M}_{J}$'s
correspond to the standard static moments\,\cite{Jackson:1975} defined
by Taylor expanding the Green's function around $\bm y=0$, a procedure
that implicitly assumes $|\bm x|\gg|\bm y|$. As indicated by Eq.\,(\ref{eq:greenexp}),
the complete expansion of the Green's function contains terms corresponding
to both $|\bm x|>|\bm y|$ as well as $|\bm y|>|\bm x|$. Thus, to
express the potentials in terms of the static nuclear moments, we
have used $\theta(x-y)=1-\theta(y-x)$ and collected all terms proportional
to $\theta(y-x)$ in the ``local'' multipole operators $\hat{\mathcal{C}}_{J}(x)$'s
and $\hat{\mathcal{M}}_{J}(x)$.\,%
\footnote{A ``local'' multipole means it can only interact with another
field when they overlap. The classical example is the extra $\delta$
function term introduced in the Cartesian multipole expansion\,\cite{Jackson:1975}.} %
Physically, the static multipoles characterize the interaction of
the electrons with a point-like nucleus, while the local multipoles
correspond to the penetration of the electrons inside the nucleus.

The multipole expansion also provides a useful framework for decomposing the interaction of the nuclear charge and current with an external potential. In the present instance we distinguish the terms in $H_{\mathrm ext}^{(\mathcal{N})} $ according to the transformation properties of the various multipole components:
\be
H_{\mathrm ext}^{(\mathcal{N})} = V_{\mathrm{ext}}^{(\mathcal{N})}+{\tilde V}_{\mathrm{ext}}^{(\tilde{\mathcal{N}})}\,,
\ee
where $V_{\mathrm{ext}}^{(\mathcal{N})}$ is even under P and T while ${\tilde V}_{\mathrm{ext}}^{(\tilde{\mathcal{N}})}$ has the opposite transformation properties. For example, $V_{\mathrm{ext}}^{(\mathcal{N})}$ contains the interaction of the nuclear magnetic dipole moment with constant applied magnetic field, while ${\tilde V}_{\mathrm{ext}}^{(\tilde{\mathcal{N}})}$ contains the interaction of the nuclear EDM with constant applied electric field, $\bm E_{0}^{(ext)}$. Substituting the potential for the latter
\be
\phi^{(ext)}(\bm y)=-\alpha\,\bm E_{0}^{(ext)}\cdot\bm y
\ee
into Eq.\,(\ref{eq:hextnuc}) leads to
\be
{\tilde V}_{\mathrm{ext}}^{(\tilde{\mathcal{N}})}=-\alpha\,{\bm d}_{\mathcal{N}}\cdot {\bm E}_0^{(ext)}\,,
\ee
where $\bm d_{\mathcal{N}}$ is the nuclear EDM given in terms of $\hat{C}_1^M$ as
\begin{align}
\bm d_{\mathcal{N}} & =\left(\frac{4\,\pi}{3}\right)^{1/2}\,\sum_{M}\, {\hat C}_{1}^{M}\,\bm e_{M}^{*}\,,
\label{eq:dndef}
\end{align}
where $\bm e_{M}$ is a spherical unit vector.

\vskip 0.1in

\noindent{\em The unperturbed and perturbed Hamiltonian}

\vskip 0.1in

We are now in a position to identify the components of unperturbed and perturbed atomic Hamiltonian. The unperturbed Hamiltonian is
\begin{equation}
\label{eq:hatom0}
H_{0}=\sum_{i=1}^{Z}\,(\beta_{i}\, m_{e}+\bm\alpha_{i}\cdot\bm p_{i})+V_{\mathrm{int}}^{(ee)}+
V_{\mathrm{int}}^{(e\mathcal{N})}+H^{{\rm nuc}}_{{\rm int}}\,.
\end{equation}
This Hamiltonian contains the following interactions:

\begin{enumerate}
\item The free electron Hamiltonian, $\beta_{i}\, m_{e}+\bm\alpha_{i}\cdot\bm p_{i}$:
The electrons are treated as relativistic, point-like particles.
\item The PCTC $e$-$e$ interaction, $V_{\mathrm{int}}^{(ee)}$: As electrons are treated relativistically,
both Coulomb and Breit interactions (the latter is not electrostatic)
are included. The scalar and vector potentials in Eq.\,(\ref{eq:pctcee})
acting on the $i$th electron in the Coulomb
gauge are
\begin{equation}
\phi_{i}^{(e)}=\sum_{j\neq i}\,\frac{1}{x_{ij}}\,,\quad\bm A_{i}^{(e)}=\sum_{j\neq i}\,\frac{1}{2\, x_{ij}}\,\left(\bm\alpha_{j}+\bm x_{ij}\frac{\bm\alpha_{j}\cdot\bm x_{ij}}{x_{ij}^{2}}\right)\,,
\end{equation}
where $x_{ij}=|\bm x_{ij}|\equiv|\bm x_{i}-\bm x_{j}|$.
\item The PCTC $e$-$\mathcal{N}$ interaction, $V_{\mathrm{int}}^{(e\mathcal{N})}$: In atomic physics,
the nucleus is typically considered as a stable, external $c$-number
source of electromagnetic fields acting on the electrons. Here, however, we take
the PCTC nuclear potentials $\phi^{(\mathcal{N})}$ and $\bm A^{(\mathcal{N})}$ [Eqs.(\ref{eq:phinuc}-\ref{eq:amult})]
to be dynamical quantities and expand
them in terms of nuclear multipole moment operators, as described above. We
will later resort to the $c$-number source approximation when considering
practical atomic calculations.
\item The PCTC internal nuclear Hamiltonian, $H^{{\rm nuc}}_{{\rm int}}$: Again,
in typical atomic computations, $H_{0}$ is diagonalized using product
wave functions that separately diagonalize $H^{{\rm nuc}}_{{\rm int}}$
and the electronic operators in Eq.\,(\ref{eq:hatom}). We will not
resort to this approximate diagonalization until Section\,\ref{sec:schiff},
so we must include $H^{{\rm nuc}}_{{\rm int}}$ explicitly.
\end{enumerate}
The perturbed Hamiltonian contains PCTC interactions between the electrons
and nucleus with the external field (no tilde) and PVTV interactions
within the atom or with the external field (terms with tildes):
\begin{align}
H_{I}= & 
\label{eq:hatomI}
V_{\mathrm{ext}}^{(e)}+V_{\mathrm{ext}}^{(\mathcal{N})}+\left[\tilde{V}_{\mathrm{ext}}^{(\tilde{e})}+\tilde{V}_{\mathrm{int}}^{(\tilde{e}\mathcal{N})}+\tilde{V}_{\mathrm{int}}^{(\tilde{e}e)}\right]+\left[\tilde{V}_{\mathrm{ext}}^{(\tilde{\mathcal{N}})}+\tilde{V}_{\mathrm{int}}^{(e\tilde{\mathcal{N}})}+{\tilde V}^{{\rm nuc}}_{{\rm int}}\right]\,,
\end{align}
where we have collected together the terms proportional to the electron EDM as well as the
terms involving PVTV nuclear couplings.
The PVTV internal nuclear potential, ${\tilde V}^{{\rm nuc}}_{{\rm int}}$, is included in the latter as one of the perturbations. Its presence will induce mixing among nuclear states of opposite parity. As discussed above, this mixing leads to non-vanishing matrix elements of the PVTV nuclear moment operators involving the ordinary nuclear charge and current operators. If ${\tilde V}^{{\rm nuc}}_{{\rm int}}$ carries a momentum dependence, the nuclear continuity equation also requires the presence of exchange charge and three-current operators,  $\hat{{\tilde\rho}}(\bm y)$ and $ \hat{\tilde{\bm j}}(\bm y)$, that have the opposite parity and time-reversal transformation properties compared to the ordinary charge and current operators. Insertion of $ {\hat{\tilde\rho}}(\bm y)$ and $ \hat{\tilde{\bm j}}(\bm y)$ into the PVTV nuclear multipole operators yields operators that have non-vanishing matrix elements between same-parity states.

\subsection{Schiff screening and its corrections\,\label{subsec:schiff}}

\begin{figure*}
\caption{Direct interactions between the external field and the electron (a) and nuclear (b)
EDMs and the induced terms involving electronic (c) and nuclear (d) polarizations (by the external field) that result in screening cancellations. In the remainder of this paper, diagram (d) is ignored.}
\label{fig:dA}

\begin{center}
\begin{tabular}{ccc}
\includegraphics[scale=0.8]{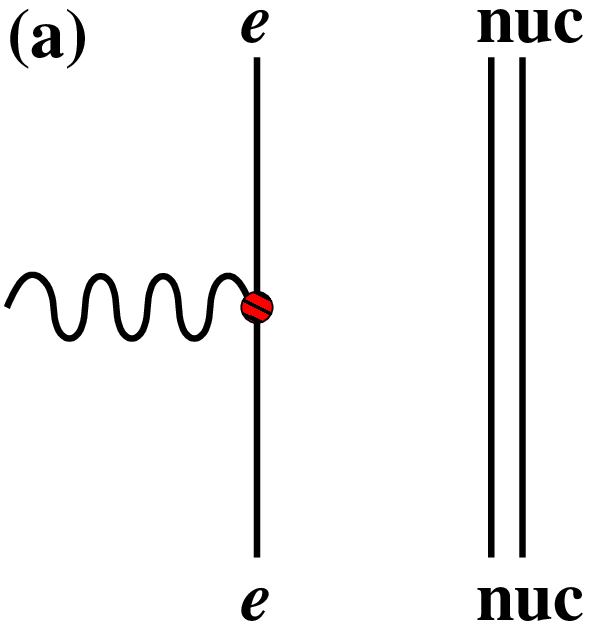}& &
\includegraphics[scale=0.8]{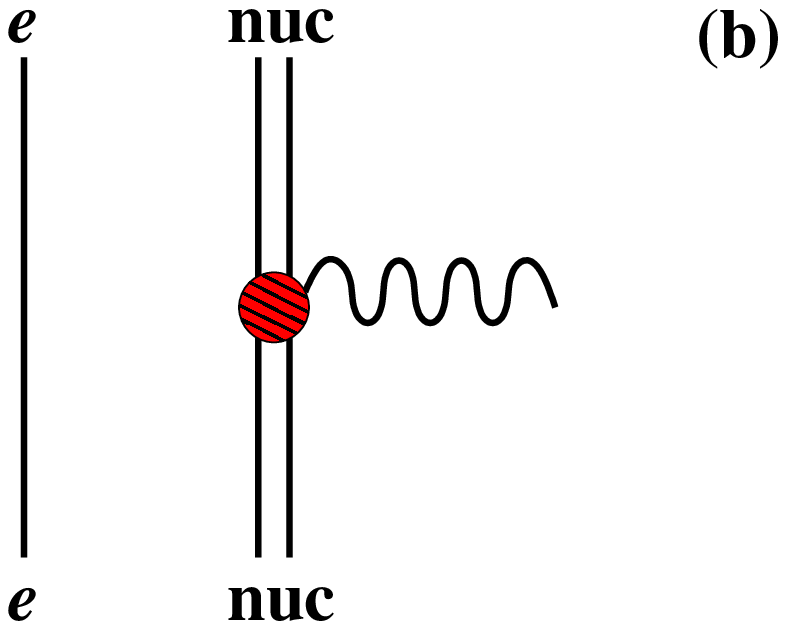}
\end{tabular}
\begin{tabular}{ccc}
\includegraphics[scale=0.8]{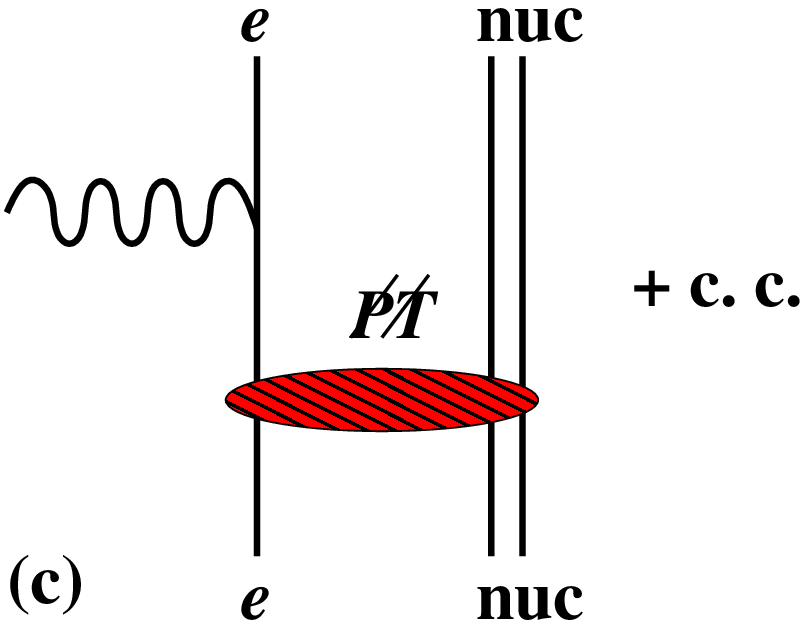}& &
\includegraphics[scale=0.8]{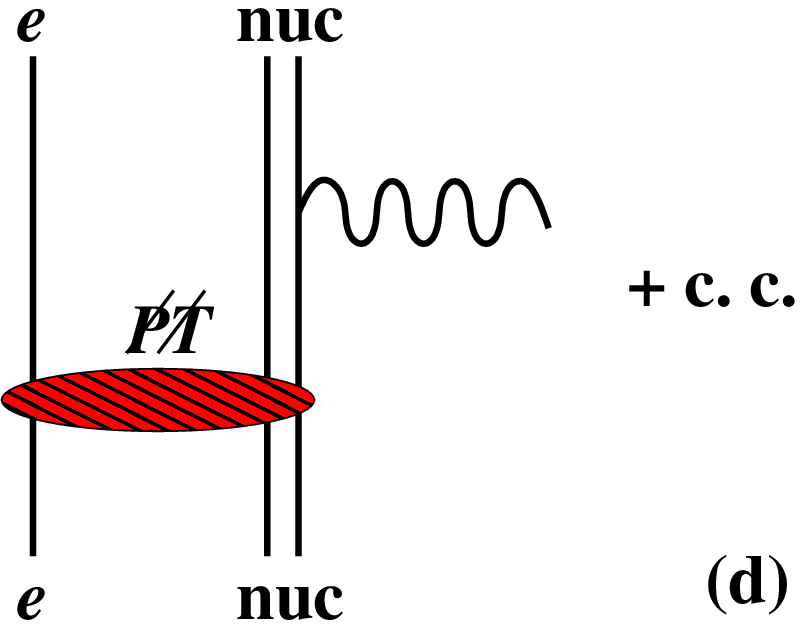}
\end{tabular}
\end{center}
\end{figure*}

The observability of the EDM of a neutral composite system is severely
hampered by the screening effect, the cancellation between direct
interactions of the external field with electron or nuclear EDMs and the
induced terms involving the internal fields induced by the applied
field.  The terms we will discuss in the next two subsections are illustrated
if Fig.\,\ref{fig:dA}. Compared to (c), (d) not only has a suppression due to the higher nuclear excitation energy: $\Delta E_{\mathrm{atom}}/\Delta E_{\mathrm{nuc}}\sim10^{-6}$, but also the $e$-$\mathcal N$ electromagnetic interaction here can only go through higher nuclear multipoles like $\mathcal C_0(x)$ and $C_2$ or $M_1$, which, in combination with the charge dipole transition matrix element, results in an additional finite-size or hyperfine suppression factor, $\sim 10^{-9}$ or $10^{-7}$, respectively. Therefore, the panel (d) can be safely ignored in the subsequent discussion.

As Schiff pointed out in deriving
his theorem\,\cite{Schiff:1963}, there exist three types of effects
that contradict the assumptions of the Schiff theorem and, thus, make
the shielding incomplete: (a) the constituent particles are relativistic;
(b) the constituents have finite size; or (c) there exist non-electrostatic
interactions between the constituents. All three effects can be present
in atomic systems: (a) the atomic electrons may be relativistic, especially
for heavy atoms; (b) the atomic nucleus has a finite spatial extent;
and (c) the $e$-$e$ and $e$-$\mathcal{N}$ electromagnetic interactions contain current-current 
components that are not electrostatic.
In this section, all of these factors will be gathered together, and
a more unified and consistent derivation of corrections to Schiff
screening effect will be presented.


In the literature one can find several ways of demonstrating that
EDMs cause no first-order energy shift, such as representing an EDM
as an infinitesimal displaced charge with the help of translational
operators as in Schiff's original paper\,\cite{Schiff:1963}, or
representing the shielding as a hypervirial theorem with the help
of the Hellmann-Feynman theorem as in Ref.\,\cite{Engel:1999np}.
In this paper, we follow the approach taken in Refs.\,\cite{Sandars:1968b,Feinberg:1977}
and rely on perturbation theory. We concentrate first on corrections
to the screening of $d_{e}$ and follow it with a treatment of the
more involved case of $d_{\mathcal{N}}$.

\subsection*{Shielding of Electron EDM}

To evaluate the consequences of shielding for an  electron EDM,
we consider the first- and second-order energy shifts
that depend on both $d_{e}$ and $\bm E^{(ext)}$:

\begin{eqnarray}
\Delta E^{(\tilde{e})}_{(1)} & = & \bra{\mathrm{g.s.}}\tilde{V}_{\mathrm{ext}}^{(\tilde{e})}\ket{\mathrm{g.s.}}\,,\label{eq:firste}\\
\Delta E^{(\tilde{e})}_{(2)} & = & \sum_{n}\,\frac{1}{E_{\mathrm{g.s.}}-E_{n}}\,\left\{ \bra{\mathrm{g.s.}}\tilde{V}_{\mathrm{int}}^{(\tilde{e}e)}+\tilde{V}_{\mathrm{int}}^{(\tilde{e}\mathcal{N})}\ket{n}\bra{n}V_{\mathrm{ext}}^{(e)}\ket{\mathrm{g.s.}}+\mathrm{c.c.}\right\} \,,
\label{eq:seconde}
\end{eqnarray}
where $\ket{\mathrm{g.s.}}$ and $\ket{n}$ denote the unperturbed
atomic (electronic plus nuclear) ground and excited states, respectively:
\begin{align}
H_{0}\ket{\mathrm{g.s.}}=E_{\mathrm{g.s.}}\ket{\mathrm{g.s}}\,, & \qquad H_{0}\ket{n}=E_{n}\ket{n}\,.
\end{align}
The first-order energy shift $\Delta E^{(\tilde{e})}_{(1)}$ arises from the interaction
of the external field with the electron EDMs, while $\Delta E^{(\tilde{e})}_{(2)}$
contains the effects due to the PCTC excitation of the atomic cloud
by the external field and its de-excitation by the PVTV $\tilde{e}$-$e$
and $\tilde{e}$-$\mathcal{N}$ interactions (and vice-versa).

Following Schiff, it is useful to express the PVTV interaction $\tilde{V}_{\mathrm{int}}^{(\tilde{e}e)}+\tilde{V}_{\mathrm{int}}^{(\tilde{e}\mathcal{N})}$
in terms of an appropriate commutator with $H_{0}$ minus correction
terms. One has\,%
\footnote{A similar commutation relation was also considered by Lindroth \emph{et
al.} in Ref.\,\cite{Lindroth:1989a} and was referred to as ``stratagem
II''. It is in contrast to the earlier work by Sandars\,\cite{Sandars:1968b}
(``stratagem I''), where a commutation relation without the $\beta$
matrix was employed. The difference between the present work and Ref.\,\cite{Lindroth:1989a}
is that neither nuclear nor external vector potentials were included
in Ref.\,\cite{Lindroth:1989a}.} %
\begin{equation}
\tilde{V}_{\mathrm{int}}^{(\tilde{e}e)}+\tilde{V}_{\mathrm{int}}^{(\tilde{e}\mathcal{N})}=-\sum_{i=1}^{Z}\,\left[d_{e}\,\beta\,\bm\sigma_{i}\cdot\bm\nabla_{i}\,,\, H_{0}\right]+2\, d_{e}\,\beta\, i\,\gamma_{5}\,\left[\bm p_{i}^{2}+\alpha\,(\bm A_{i}^{(\mathcal{N})}-\nicefrac{1}{2}\,\bm A_{i}^{(e)})\cdot\bm p_{i}\right]\,.\label{eq:de-closure}
\end{equation}
In the non-relativistic limit, the first term on the RHS of Eq.\,(\ref{eq:de-closure})
contains the commutator of $H_{0}$ with Schiff's displacement operator,
$d_{e}\,\bm\sigma\cdot\bm\nabla$. Letting $H_{0}$ act on either
side of the relativistic form of this operator in Eq.\,(\ref{eq:seconde})
leads to the energy difference $E_{\mathrm{g.s.}}-E_{n}$ that cancels
the corresponding energy denominator. One may then carry out the closure
sum on the states $\ket{n}$, leading to
\begin{eqnarray}
\Delta E^{(\tilde{e})}_{(2)} & = & \sum_{i=1}^{Z}\,\bra{\mathrm{g.s.}}\left[d_{e}\,\beta\,\bm\sigma_{i}\cdot\bm\nabla_{i}\,,V_{\mathrm{ext}}^{(e)}\right]\ket{\mathrm{g.s.}}\nonumber \\
&  & +2\, i\, d_{e}\,\sum_{n}\frac{1}{E_{\mathrm{g.s.}}-E_{n}}\,\Bigg\{\bra{\mathrm{g.s.}}\sum_{i=1}^{Z}\,\beta\,\gamma_{5}\,\left[\bm p_{i}^{2}+\alpha\,(\bm A_{i}^{(\mathcal{N})}-\nicefrac{1}{2}\,\bm A_{i}^{(e)})\cdot\bm p_{i}\right]\ket{n}\nonumber \\
&  & \times\bra{n}{V}_{\mathrm{ext}}^{(e)}\ket{\mathrm{g.s.}}+\mathrm{c.c.}\,\Bigg\}\,.\label{eq:de-closure2}
\end{eqnarray}
After performing some Dirac algebra on the first term on the RHS
of Eq.\,(\ref{eq:de-closure2}) and adding $\Delta E^{(\tilde{e})}_{(1)}$ and
$\Delta E^{(\tilde{e})}_{(2)}$ we obtain
\begin{align}
\Delta E^{(\tilde{e})}_{(1)}+\Delta E^{(\tilde{e})}_{(2)} & =\Delta E_{\mathrm{(1')}}^{(\tilde{e})}+\Delta E_{\mathrm{(2')}}^{(\tilde{e})}\,,
\end{align}
where
\begin{eqnarray}
\Delta E_{\mathrm{(1')}}^{(\tilde{e})} & = & \bra{\mathrm{g.s.}}\sum_{i=1}^{Z}\,2\,\alpha\, d_{e}\,\beta\, i\,\gamma_{5}\,\bm A_{i}^{(ext)}\cdot\bm p_{i}\ket{\mathrm{g.s.}}\,,
\label{eq:DE-e1-scr}\\
\Delta E_{\mathrm{(2')}}^{(\tilde{e})} & = & 2\, i\, d_{e}\,\sum_{n}\,\frac{1}{E_{\mathrm{g.s.}}-E_{n}}\,\bigg\{\bra{\mathrm{g.s.}}\sum_{i=1}^{Z}\,\beta\,\gamma_{5}\,\left[\bm p_{i}^{2}+\alpha\,(\bm A_{i}^{(\mathcal{N})}-\nicefrac{1}{2}\,\bm A_{i}^{(e)})\cdot\bm p_{i}\right]\ket{n}\nonumber \\
&  & \times\bra{n}{V}_{\mathrm{ext}}^{(e)}\ket{\mathrm{g.s.}}+\mathrm{c.c.}\bigg\}\,.
\label{eq:DE-e2-scr}
\end{eqnarray}
The effective residual first-order contribution $\Delta E_{\mathrm{(1')}}^{(\tilde{e})}$ arises from a
partial cancellation between $\Delta E^{(\tilde{e})}_{(1)}$ and the commutator
term of $\Delta E^{(\tilde{e})}_{(2)}$ in Eq.\,(\ref{eq:de-closure2}). Both $\Delta E^{(\tilde{e})}_{(1')}$ and
$\Delta E^{(\tilde{e})}_{(2')}$ involve matrix elements of $\gamma_{5}$ that connects large
(upper) and small (lower) components of Dirac wave functions. Thus,
for non-relativistic (NR) electrons (with vanishing lower components),
$\Delta E_{\mathrm{(1')}}^{(\tilde{e})}=\Delta E_{\mathrm{(2')}}^{(\tilde{e})}=0$,
implying complete shielding of electron EDMs.

It should be noted that the procedure for using a closure sum to
evaluate the effects of shielding is not unique. The first discussion of relativistic
effects by Sandars\,\cite{Sandars:1968b} relies on expressing $\beta$
matrix in the EDM interaction as $1+(\beta-1)$, and using closure
just to remove the ``$1$'' term while leaving the $(\beta-1)$
term intact. This approach leads to a less complete cancellation of
$\Delta E^{(\tilde{e})}_{(1)}$. However, as the use of $(\beta-1)$ also requires
nonvanishing small components, it has the same NR limit as the $\gamma_{5}$
formalism above. In a later publication\,\cite{Lindroth:1989a},
Lindroth \emph{et al.} argued that although the $\gamma_{5}$ formalism
has the advantage of being purely one-body when no Breit interaction
is present, it is not formally of order $Z^{2}\,\alpha^{2}$ in contrast
to the $(\beta-1)$ formalism. This might introduce complications
when complex atomic many-body calculations intend to reach this level
of accuracy\,\cite{Lindroth:1989a}. However, no detailed calculation
further substantiates this claim. In this paper, the $\gamma_{5}$
formalism is chosen because it is simpler.

\subsection*{Shielding of the Nuclear EDM}

To evaluate the effects of shielding of a nuclear EDM,
we follow a procedure similar to that used in the case of the electron
and consider energy shifts that are linear in $\bm d_{\mathcal{N}}$
and the external field to the second order in perturbation theory. However, we
depart from the previous notation somewhat in that we make the polarization effect of
$\tilde{V}^{\mathrm{nuc}}_{\mathrm{int}}$ explicit in the nuclear wave functions.
This yields
\begin{eqnarray}
\label{eq:nuc1}
\Delta E^{(\tilde{\mathcal{N}})}_{(1)} &=& \bra{\mathrm{g.s._{\tilde{\mathcal N}}}}\tilde{V}_{\mathrm{ext}}^{(\tilde{\mathcal{N}})}\ket{\mathrm{g.s.}_{\tilde{\mathcal N}}}\,,\\
\label{eq:nuc2}
\Delta E^{(\tilde{\mathcal{N}})}_{(2)} &=& \sum_{n}\,\frac{1}{E_{\mathrm{g.s.}}-E_{n}}\,\left\{ \bra{\mathrm{g.s.}}V_{\mathrm{ext}}^{(e)}\ket{n}\bra{n_{\tilde{\mathcal N}}}\tilde{V}_{\mathrm{int}}^{(e\tilde{\mathcal{N}})}\ket{\mathrm{g.s.}_{\tilde{\mathcal N}}}+\mathrm{c.c.}\right\} \,,
\end{eqnarray}
For example, if the atomic wave function has the direct-product form  $\ket{n} = \ket{n_{\mathcal{N}}}\otimes\ket{n_e}$, then
\be
\ket{n_{\tilde{\mathcal N}}}= \sum_{m}\,\frac{1}{E_{n_{\mathcal{N}}}-E_{m_{\mathcal{N}}}} \,_{\mathcal{N}}\bra{m} \tilde{V}^{\mathrm{nuc}}_{\mathrm{int}} \ket{n}_{\mathcal{N}} \ket{m}_{\mathcal{N}} \otimes \ket{n}_e \equiv \ket{\tilde{n}}_{\mathcal{N}} \otimes \ket{n}_e\,.
\label{eq:pol-wf}
\ee
It is understood that PVTV effects are to be evaluated only in first order.  Thus the meaning of the notation
in Eqs.\,(\ref{eq:nuc1}) and (\ref{eq:nuc2}) is:
\begin{itemize}
\item Contributions involving PVTV charges or three-currents in $\tilde{V}_{\mathrm{ext}}^{(\tilde{\mathcal{N}})}$
or $\tilde{V}_{\mathrm{int}}^{(e \tilde{\mathcal{N}})}$ are to be evaluated with unperturbed wave functions.
\item Contributions involving $\tilde{V}^{\mathrm{nuc}}_{\mathrm{int}}$ correspond to matrix 
elements with PCTC charges or currents in  $\tilde{V}_{\mathrm{ext}}^{(\tilde{\mathcal{N}})}$
or $\tilde{V}_{\mathrm{int}}^{(e \tilde{\mathcal{N}})}$.  Furthermore, $\tilde{V}^{\mathrm{nuc}}_{\mathrm{int}}$
is to be treated in first order, modifying the bra or the ket state, but not both.
\end{itemize}
Below we suppress the explicit subscript $\tilde{\mathcal{N}}$ on bra and ket states involving operators coupling to the
nucleus, but implicitly $\tilde{V}^{\mathrm{nuc}}_{\mathrm{int}}$ has been absorbed into nuclear wave functions and its effects will be retained to first order in the
PVTV.

To proceed, we first write $\bm d_{\mathcal{N}}$ in a spherical basis using Eq.\,(\ref{eq:dndef})
and observe that
\begin{align}
\left[\bm d_{\mathcal{N}}\cdot\bm\nabla\,,\,\frac{1}{x}\right] & =-\left(\frac{4\,\pi}{3}\right)\,\hat{C}_{1}\otimes Y_{1}(\Omega_{x})\,\frac{1}{x^{2}}\,.
\end{align}
Letting
\begin{align}
\Delta H_{0} & \equiv H_{0}-H_{0}^{e\textrm{--}\mathcal{N}}(C_{0}\textrm{ only})=H_{0}+Z\,\alpha\,\sum_{i=1}^{Z}\,\frac{1}{x_{i}}
\end{align}
be the unperturbed Hamiltonian without the static electron-nucleus
PCTC monopole Coulomb interaction, we have that
\begin{eqnarray}
-\alpha\,\tilde{\phi}_{i}^{(C_{1})} & = & -\alpha\,\left(\frac{4\pi}{3}\right)\,\hat{C}_{1}\odot\sum_{i=1}^{Z}Y_{1}(\Omega_{x_{i}})\,\frac{1}{x_{i}^{2}}\nonumber \\
& = & \alpha\sum_{i=1}^{Z}\,\left[\bm d_{\mathcal{N}}\cdot\bm\nabla_{i}\,,\,\sum_{j=1}^{Z}\,\frac{1}{x_{j}}\right]=-\sum_{i=1}^{Z}\,\left[\frac{\bm d_{\mathcal{N}}}{Z}\cdot\bm\nabla_{i}\,,\, H_{0}-\Delta H_{0}\right]\,.
\label{eq:dN-closure1}
\end{eqnarray}
Substituting Eq.\,(\ref{eq:dN-closure1}) into $\Delta E^{(\tilde{\mathcal{N}})}_{(2)}$
and letting $H_{0}$ act on either side of the displacement operator leads
to the energy difference $E_{{\rm g.s.}}-E_{n}$ that cancels the
energy denominator before and allows us to carry out a closure sum
for this term. Doing so and letting
\begin{align}
\Delta\tilde{V}_{\mathrm{int}}^{(e\tilde{\mathcal{N}})} & =\tilde{V}_{\mathrm{int}}^{(e\tilde{\mathcal{N}})}+\alpha\,\tilde{\phi}_{i}^{(C_{1})}
\end{align}
leads to
\begin{eqnarray}
\Delta E^{(\tilde{\mathcal{N}})}_{(2)} & = & \sum_{n}\,\frac{1}{E_{\mathrm{g.s.}}-E_{n}}\,\left\{ \bra{\mathrm{g.s.}}V_{\mathrm{ext}}^{(e)}\ket{n}\bra{n}-\alpha\,\tilde{\phi}_{i}^{(C_{1})}+\Delta\tilde{V}_{{\rm int}}^{(e\tilde{\mathcal{N}})}\ket{\mathrm{g.s.}}+\mathrm{c.c.}\right\} \nonumber \\
& = & \frac{1}{Z}\,\bra{\mathrm{g.s.}}\left[\sum_{i=1}^{Z}\,\bm d_{\mathcal{N}}\cdot\bm\nabla_{i}\,,\, V_{\mathrm{ext}}^{(e)}\right]\ket{\mathrm{g.s.}}+\sum_{n}\,\frac{1}{E_{\mathrm{g.s.}}-E_{n}}\nonumber \\
&  & \times\left\{ \bra{\mathrm{g.s.}}V_{\mathrm{ext}}^{(e)}\ket{n}\bra{n}\Delta\tilde{V}_{{\rm int}}^{(e\tilde{\mathcal{N}})}+\sum_{i=1}^{Z}\,\left[\frac{\bm d_{\mathcal{N}}}{Z}\cdot\bm\nabla_{i}\,,\,\Delta H_{0}\right]\ket{\mathrm{g.s.}}+\mathrm{c.c.}\right\}\,.
\end{eqnarray}

Now observe that
\begin{eqnarray}
\frac{1}{Z}\,\left[\sum_{i=1}^{Z}\,\bm d_{\mathcal{N}}\cdot\bm\nabla_{i}\,,\, V_{\mathrm{ext}}^{(e)}\right] & = & -\frac{\alpha}{Z}\,\sum_{i,j=1}^{Z}\,\left[\bm d_{\mathcal{N}}\cdot\bm\nabla_{i}\,,\,\phi_{j}^{(ext)}-\bm\alpha_{j}\cdot\bm A_{j}^{(ext)}\right]\nonumber \\ 
& = & \alpha\,\bm d_{\mathcal{N}}\cdot\bm E_{0}^{(ext)}+\frac{\alpha}{Z}\,\sum_{i=1}^{Z}\,\left[\bm d_{\mathcal{N}}\cdot\bm\nabla_{i}\,,\,\bm\alpha_{i}\cdot\bm A_{i}^{(ext)}\right]\,,
\end{eqnarray}
where we have used the electrostatic stability condition $Z\,\bm E_{0}^{(ext)}-\sum_{i}\,\bm E_{i}^{(ext)}=0$.
Adding the first- and second-order energy shifts leads to
\begin{align}
\Delta E^{(\tilde{\mathcal{N}})}_{(1)}+\Delta E^{(\tilde{\mathcal{N}})}_{(2)}= & \frac{\alpha}{Z}\,\bra{\mathrm{g.s.}}\sum_{i=1}^{Z}\,\left[\bm d_{\mathcal{N}}\cdot\bm\nabla_{i}\,,\,\bm\alpha_{i}\cdot\bm A_{i}^{(ext)}\right]\ket{\mathrm{g.s.}}+\sum_{n}\,\frac{1}{E_{\mathrm{g.s.}}-E_{n}}\nonumber \\
& \times\left\{ \bra{\mathrm{g.s.}}V_{\mathrm{ext}}^{(e)}\ket{n}\bra{n}\Delta\tilde{V}_{{\rm int}}^{(e\tilde{\mathcal{N}})}+\sum_{i=1}^{Z}\,\left[\frac{\bm d_{\mathcal{N}}}{Z}\cdot\bm\nabla_{i}\,,\,\Delta H_{0}\right]\ket{\mathrm{g.s.}}+\mathrm{c.c.}\right\} \,.
\label{eq:dN-final}
\end{align}

Equation\,(\ref{eq:dN-final}) demonstrates the cancellation of all
terms proportional to $\bm d_{\mathcal{N}}\cdot\bm E_{0}^{(ext)}$
due to the screening effect and allows us to identify systematically
all corrections to this screening. We elaborate on these corrections,
including the correction associated with the Schiff moment, below.

\subsection*{Shielding Corrections}

From Eqs.\,(\ref{eq:DE-e1-scr}, \ref{eq:DE-e2-scr}) and (\ref{eq:dN-final}),
one may determine all the corrections to Schiff screening that are
linear in $d_{e}$ or $d_{\mathcal{N}}$ and that occur to the second
order in perturbation theory. For purposes of future discussion, it
is useful to classify the corrections as follows:

\vskip 0.1in

\noindent \emph{Ground state matrix elements}. The following corrections
arise solely from the interaction with the external vector potential
$\bm A_{i}^{(ext)}$:
\begin{align}
\Delta E_{(1)} & =\bra{\mathrm{g.s.}}\hat{O}_{\mathrm{ext}}^{(\tilde{e})}+\hat{O}_{\mathrm{ext}}^{(\tilde{\mathcal{N}})}\ket{\mathrm{g.s.}}\,,
\end{align}
where
\begin{eqnarray}
\hat{O}_{{\rm ext}}^{(\tilde{e})} & = & \sum_{i=1}^{Z}\,2\,\alpha\, d_{e}\,\beta\, i\,\gamma_{5}\,\bm A_{i}^{(ext)}\cdot\bm p_{i}\,,\\
\hat{O}_{{\rm ext}}^{(\tilde{\mathcal{N}})} & = & \frac{\alpha}{Z}\,\sum_{i=1}^{Z}\,\left[\bm d_{\mathcal{N}}\cdot\bm\nabla_{i}\,,\,\bm\alpha_{i}\cdot\bm A_{i}^{(ext)}\right]\,.
\end{eqnarray}

\noindent \emph{Internal excitations}. The remaining corrections involve
excitation of atomic states by $V_{\mathrm{ext}}^{(e)}$ and de-excitation
by operators proportional to $d_{e}$, $d_{\mathcal{N}}$, or higher
PVTV nuclear moments. The corresponding energy shift is
\begin{align}
\Delta E_{(2)} & =\sum_{n}\,\frac{1}{E_{\mathrm{g.s.}}-E_{n}}\left\{ \bra{\mathrm{g.s.}}V_{\mathrm{ext}}^{(e)}\ket{n}\bra{n}\hat{O}_{\mathrm{int}}^{(\tilde{e})}+\sum_{i=1}^{Z}\,\sum_{k=1}^{9}\,\hat{O}_{k,i}^{(\tilde{\mathcal{N}})}\ket{\mathrm{g.s.}}+\mathrm{c.c.}\right\} \,,
\label{eq:deltaE2screen}
\end{align}
where
\begin{align}
\hat{O}_{\mathrm{int}}^{(\tilde{e})} & =2\, i\, d_{e}\,\sum_{i=1}^{Z}\,\beta\,\gamma_{5}\,\left[\bm p_{i}^{2}+\alpha\,\left(\bm A_{i}^{(\mathcal{N})}-\nicefrac{1}{2}\,\bm A_{i}^{(e)}\right)\cdot\bm p_{i}\right]\,,
\end{align}
and where, with all electronic subscripts ``$i$'' suppressed,
\begin{eqnarray}
\hat{O}_{1}^{(\tilde{\mathcal{N}})} & = & \frac{4\pi\alpha}{Z}\,\sum_{J\ge2}^{\mathrm{even}}\,\frac{1}{x^{J+2}}\Bigg\{\sqrt{\frac{J+1}{2\, J+3}}\, Y_{J+1}(\hat{x})\odot C_{J}\nonumber \\
&  & +\frac{1}{2\, J+1}\,(x\, Y_{J}(\hat{x})\,\overleftrightarrow{\bm\nabla})\dot{\odot}[\bm d_{\mathcal{N}}\,,\, C_{J}]\Bigg\}\,,\label{eq:O1}\\
\hat{O}_{2}^{(\tilde{\mathcal{N}})} & = & \frac{4\pi\alpha}{Z}\,\sum_{J\ge2}^{\mathrm{even}}\,\frac{1}{x^{J+2}}\Bigg\{\sqrt{\frac{J}{2\, J-1}}\, Y_{J-1}(\hat{x})\odot[\bm d_{\mathcal{N}}\otimes\mathcal{C}_{J<}(x)]_{J-1}\nonumber \\
&  & -\sqrt{\frac{J+1}{2\, J+3}}\, Y_{J+1}(\hat{x})\odot[\bm d_{\mathcal{N}}\otimes\mathcal{C}_{J>}(x)]_{J+1}+\frac{1}{2\, J+1}\,(x\, Y_{J}(\hat{x})\,\overleftrightarrow{\bm\nabla})\dot{\odot}[\bm d_{\mathcal{N}}\,,\,\mathcal{C}_{J}(x)]\Bigg\}\nonumber \\
&  & -\frac{4\,\pi\,\alpha}{Z\, x^{2}}\,\sqrt{\frac{1}{3}}\, Y_{1}(\hat{x})\odot[\bm d_{\mathcal{N}}\otimes\mathcal{C}_{0>}(x)]_{1}+\frac{4\,\pi\,\alpha}{Z\, x^{2}}\,(x\, Y_{0}(\hat{x})\,\overleftrightarrow{\bm\nabla})\dot{\odot}[\bm d_{\mathcal{N}}\,,\,\mathcal{C}_{0}(x)]\,,\\
\hat{O}_{3}^{(\tilde{\mathcal{N}})} & = & \frac{4\pi\alpha}{Z}\,\sum_{J\ge1}^{\mathrm{odd}}\,\frac{1}{x^{J+2}}\Bigg\{-\sqrt{\frac{1}{(J+1)\,(2\, J+1)}}\,[Y_{J+1}(\hat{x})\otimes\bm\alpha]_{J}\odot[\bm d_{\mathcal{N}}\otimes M_{J}]_{J}\nonumber \\
&  & +\sqrt{\frac{J\,(J+2)}{(J+1)\,(2\, J+3)}}\,[Y_{J+1}(\hat{x})\otimes\bm\alpha]_{J+1}\odot[\bm d_{\mathcal{N}}\otimes M_{J}]_{J+1}\nonumber \\
&  & -\frac{1}{2\, J+1}\,(x\,[Y_{J}(\hat{x})\otimes\bm\alpha]_{J}\,\overleftrightarrow{\bm\nabla})\dot{\odot}[\bm d_{\mathcal{N}}\,,\, M_{J}]\Bigg\}\,,\\
\hat{O}_{4}^{(\tilde{\mathcal{N}})} & = & \frac{4\pi\alpha}{Z}\,\sum_{J\ge1}^{\mathrm{odd}}\,\frac{1}{x^{J+2}}\Bigg\{\sqrt{\frac{1}{(J+1)\,(2\, J+1)}}\,[Y_{J+1}(\hat{x})\otimes\bm\alpha]_{J}\odot[\bm d_{\mathcal{N}}\otimes\mathcal{M}_{J>}(x)]_{J}\nonumber \\
&  & +\sqrt{\frac{J\,(J+2)}{(J+1)\,(2\, J+3)}}\,[Y_{J+1}(\hat{x})\otimes\bm\alpha]_{J+1}\odot[\bm d_{\mathcal{N}}\otimes\mathcal{M}_{J>}(x)]_{J+1}\nonumber \\
&  & +\sqrt{\frac{(J+1)\,(J+3)}{(J+2)\,(2\, J+3)}}\,[Y_{J+1}(\hat{x})\otimes\bm\alpha]_{J+1}\odot[\bm d_{\mathcal{N}}\otimes\mathcal{M}_{J<}(x)]_{J+1}\nonumber \\
&  & +\sqrt{\frac{1}{J\,(2\, J+1)}}\,[Y_{J-1}(\hat{x})\otimes\bm\alpha]_{J}\odot[\bm d_{\mathcal{N}}\otimes\mathcal{M}_{J<}(x)]_{J}\nonumber \\
&  & -\frac{1}{2\, J+1}\,(x\,[Y_{J}(\hat{x})\otimes\bm\alpha]_{J}\,\overleftrightarrow{\bm\nabla})\odot[\bm d_{\mathcal{N}}\,,\,\mathcal{M}_{J}(x)]\Bigg\}\,,\\
\hat{O}_{5}^{(\tilde{\mathcal{N}})} & = & \left[\frac{\bm d_{\mathcal{N}}}{Z}\cdot\overleftrightarrow{\bm\nabla}\,,\, H^{{\rm nuc}}_{{\rm int}}\right]\,,\\
\hat{O}_{6}^{(\tilde{\mathcal{N}})} & = & -\sum_{J\ge3}^{\mathrm{odd}}\,\frac{4\,\pi\,\alpha}{x^{J+1}}\,\frac{1}{2\, J+1}\, Y_{J}(\hat{x})\odot C_{J}\,,\\
\hat{O}_{7}^{(\tilde{\mathcal{N}})} & = & -\sum_{J\ge1}^{\mathrm{odd}}\,\frac{4\,\pi\,\alpha}{x^{J+1}}\,\frac{1}{2\, J+1}\, Y_{J}(\hat{x})\odot\mathcal{C}_{J}(x)\,,\\
\hat{O}_{8}^{(\tilde{\mathcal{N}})} & = & -\sum_{J\ge2}^{\mathrm{even}}\,\frac{4\,\pi\,\alpha}{x^{J+1}}\,\frac{1}{2\, J+1}\,[Y_{J}(\hat{x})\otimes\bm\alpha]_{J}\odot M_{J}\,,\\
\hat{O}_{9}^{(\tilde{\mathcal{N}})} & = & -\sum_{J\ge2}^{\mathrm{even}}\,\frac{4\,\pi\,\alpha}{x^{J+1}}\,\frac{1}{2\, J+1}\,[Y_{J}(\hat{x})\otimes\bm\alpha]_{J}\odot\mathcal{M}_{J}(x)\,.
\label{eq:O9}
\end{eqnarray}
We refer to Appendix\,\ref{sec:symmetrization} for more details about the derivation of
Eqs.\,(\ref{eq:O1}--\ref{eq:O9}) from Eq.\,(\ref{eq:dN-final}) and note important conventions here:

\begin{enumerate}
\item The nuclear composite operators involving $\bm{d}_{\mathcal{N}}$ and any
PCTC nuclear multipole operator $X_{j}$ should be realized as
\begin{equation}
[\bm d_{\mathcal{N}}\otimes X_{j}]_{J}^{M}=[\bm d_{\mathcal{N}}\otimes X_{j}]_{J}^{\mathrm{(sym)}M}\equiv\sum_{\lambda,m}\braket{1\lambda,jm|JM}\{\bm d_{\mathcal{N}}^{\lambda}\,,\, X_{j}^{m}\}/2\,.
\label{eq:sym_n}
\end{equation}
As two nuclear density operators not necessarily commute, the part
which does not commute is treated explicitly in the terms in Eqs.\,(\ref{eq:O1}--\ref{eq:O9}) involving the commutator of $\bm d_{\mathcal{N}}$ with other nuclear operators.
\item The special gradient operator $\overleftrightarrow{\bm\nabla}$ only
acts on the electronic bra and ket states as
\begin{align}
_{e}\braket{\,|\mathcal{O}({\bm x})\,\overleftrightarrow{\bm\nabla}|\, }_{e} & \equiv
{}_{e}\braket{\, |\{\mathcal{O}({\bm x})\,,\,\bm\nabla\}|\, }_{e}/2=({}_{e}\braket{\, |\mathcal{O}({\bm x})\,\overrightarrow{\nabla}|\, }_{e}-{}_{e}\braket{\, |\overleftarrow{\nabla}\, \mathcal{O}({\bm x})|\, }_{e})/2\,,
\label{eq:sym_e}
\end{align}
where $\mathcal{O}({\bm x})$ denotes a generic operator that depends on electronic degrees of freedom and where in the second equality we have performed an integration by parts.

\item The operator ``$\dot{\odot}$'' denotes a double scalar product
as
\begin{align}
\left(\mathcal{O}_{J}(\bm x)\,\overleftrightarrow{\bm\nabla}\right)\dot{\odot}\left[\bm d_{\mathcal{N}}\,,\, X_{J}\right] & =\sum_{\lambda,M}\,\left(\mathcal{O}_{J}^{M*}(\bm x)\,\overleftrightarrow{\bm\nabla}^{\lambda*}\right)\,\left[\bm d_{\mathcal{N}}^{\lambda}\,,\, X_{J}^{M}\right]\,.
\label{eq:dbldot}
\end{align}

\end{enumerate}
We make several observations about the list of operators $\hat{O}_{k}^{(\tilde{\mathcal{N}})}$.

\begin{itemize}
\item [(i)] The operators $\hat{O}_{1-5}^{(\tilde{\mathcal{N}})}$ are generated
by the commutator of the displacement operator $\bm d_{\mathcal{N}}\cdot\bm\nabla$
with the non-$C_{0}$ part of the unperturbed Hamiltonian, $\Delta H_{0}$.
\item [(ii)] Physically, operators $\hat{O}_{1,3}^{(\tilde{\mathcal{N}})}$ correspond
to ``displacement'' of the static PCTC Coulomb and magnetic electron-nucleus
interactions due to the rearrangement of the atomic electrons needed
to maintain electrostatic equilibrium in the presence of the external
field. The operators $\hat{O}_{2,4}^{(\tilde{\mathcal{N}})}$ describe the
corresponding effects of displacing the PCTC penetration (non-static)
$e$-$\mathcal{N}$ multipole interactions.
\item [(iii)] Operator $\hat{O}_{5}^{(\tilde{\mathcal{N}})}$ describes the response
of the internal nuclear degrees of freedom to the external field,
again as needed to maintain electrostatic equilibrium.
\item [(iv)] Operators $\hat{O}_{6-9}^{(\tilde{\mathcal{N}})}$ characterize the
effects of the ``local'' EDM (i.e., $\mathcal{C}_{1}$), PVTV
magnetic and higher ($J\geq3$) PVTV charge multipole interactions, including both the static multipole interactions, $\hat{O}_{6,8}^{(\tilde{\mathcal{N}})}$, and penetration terms, $\hat{O}_{7,9}^{(\tilde{\mathcal{N}})}$.
\item [(v)] The leading, non-magnetic correction to Schiff screening due
to the finite spatial extent and internal structure of the nucleus
is the Schiff moment. It therefore arises from the penetration or
local multipoles $\mathcal{C}_{J\,>,<}(x)$ that appear with a $Y_{1}(\hat{x})$;
a gradient acting on the electronic coordinates coupled to a $Y_{0}(\hat{x})$
or $Y_{2}(\hat{x})$; and the internal nuclear interaction in conjunction
with a gradient acting on the electronic coordinates. Such terms
appear in $\hat{O}_{1,2,5,7}^{(\tilde{\mathcal{N}})}$ . In the following
section, we assemble these terms into the Schiff moment operator and
compare with the form of the operator previously used in the literature.

\end{itemize}

\section{The Schiff moment operator\,\label{sec:schiff}}

To arrive at the Schiff moment operator, we collect all terms in $\hat{O}_{1,2,5,7}^{(\tilde{\mathcal{N}})}$ that are proportional to $Y_{1}(\hat{x})$ or that contain terms with
a gradient acting on the atomic coordinate and transform like a spherical
vector:
\begin{eqnarray}
\left[\hat{O}_{1,2,5,7}^{(\tilde{\mathcal{N}})}\right]_{Y_{1}} & = & -\frac{4\,\pi\,\alpha}{Z\, x^{2}}\,\Bigg\{\sqrt{\frac{1}{3}}\, Y_{1}(\hat{x})\odot\left[\bm d_{\mathcal{N}}\otimes\left(\mathcal{C}_{0>}(x)-\frac{\sqrt{2}}{x^{2}}\,\mathcal{C}_{2<}(x)\right)\right]_{1}+Z\, Y_{1}(\hat{x})\odot\frac{1}{3}\,\mathcal{C}_{1}(x)\nonumber \\
&  & -(x\, Y_{0}(\hat{x})\,\overleftrightarrow{\bm\nabla})\dot{\odot}\left[\bm d_{\mathcal{N}}\,,\,\mathcal{C}_{0}(x)\right]-(x\, Y_{2}(\hat{x})\,\overleftrightarrow{\bm\nabla})\dot{\odot}\left[\bm d_{\mathcal{N}}\,,\,\frac{1}{5\, x^{2}}\,(C_{2}+\mathcal{C}_{2}(x))\right]\Bigg\}\nonumber \\
&  & +\left[\frac{\bm d_{\mathcal{N}}}{Z}\cdot\overleftrightarrow{\bm\nabla}\,,\, H^{{\rm nuc}}_{{\rm int}}\right]\ \ \equiv\ \ \hat{O}_{{\rm Schiff}}^{{\rm atomic}}\,. \label{eq:Schiff-E1a1}
\end{eqnarray}
Note that the terms containing the gradient $\bm\nabla$ acting on
the electronic coordinate will connect electronic states differing
by one unit of orbital angular momentum, as can be seen by performing
an integration by parts and letting $\bm\nabla$ act on the initial
and final electronic wave functions. Consequently, we include these
operators in $\hat{O}_{{\rm Schiff}}^{{\rm atomic}}$.

The operator $\hat{O}_{{\rm Schiff}}^{{\rm atomic}}$ generates the
leading contributions to an atomic EDM that depend only on the nuclear
charge density operator and not on the current density operator. The
form given in Eq.\,(\ref{eq:Schiff-E1a1}) is general and could,
in principle, be used to compute the atomic EDM generated by P- and
T-odd interactions in the nucleus. From Eq.\,(\ref{eq:deltaE2screen}),
we observe that $\hat{O}_{{\rm Schiff}}^{{\rm atomic}}$ will induce
mixing of opposite parity states into the atomic ground state, thereby
allowing for a shift in the atomic energy when a constant external
electric field is applied, \textit{i.e.}, $V_{{\rm ext}}^{(e)}$.
In practice, the typical atomic ground state is taken by be a direct
product of the nuclear and electronic ground states, as in Eq.\,(\ref{eq:fact}):
\begin{align}
\ket{\mathrm{g.s.}} & =\ket{\mathrm{g.s.}}_{\mathcal{N}}\otimes\ket{\mathrm{g.s.}}_{e}\,.
\end{align}

It is particularly interesting to consider systems in which the electronic
ground state is a S-state ($L=0$) that has a comparatively larger penetration probability due the lack of the centrifugal barrier. Since $\hat{O}_{{\rm Schiff}}^{{\rm atomic}}$
transforms as a rank-one tensor in the space of electronic coordinates,
it will mix P states into the electronic ground state (relevant
for many, but not all atomic EDM experiments). Thus, it is useful
to derive a form for this operator applicable to this situation. In
doing so, we can derive a nuclear Schiff moment operator that corresponds
to -- but differs in form from -- the Schiff moment operator used
elsewhere in the literature. To that end, we begin by considering
the first three terms in $\hat{O}_{{\rm Schiff}}^{{\rm atomic}}$
and compute their S-P electronic matrix elements.

The electronic wave functions for the S and P states are:
\begin{align}
\psi_{S}(\bm x) & =\braket{x|S,m}=u_{S}(x)\, Y_{0}(\hat{x})\,\chi_{1/2}^{m}\,,\\
\psi_{P}(\bm x) & =\braket{x|P,\lambda\, m^{'}}=u_{P}(x)\, Y_{1}^{\lambda}(\hat{x})\,\chi_{1/2}^{m^{'}}\,,
\end{align}
where $\lambda$ is the magnetic projection of the P state ($\lambda=0$ for the S state), and
$\chi_{1/2}^{m}$ is the spin wave function. We will henceforth
neglect the spin degrees of freedom as they are not relevant to the
derivation of the effective nuclear operator -- $m$ should be a conserved
quantum number for this case. In order to arrive at the effective nuclear operator, we consider the following polynomial expansions of electronic radial wave functions near the origin:
\begin{align}
u_{S}(x) & =\sum_{k\ge0}\, a_{k}\, x^{k}\,,\\
u_{P}(x) & =\sum_{k\ge1}\, b_{k}\, x^{k}\,.
\end{align}

We now consider the matrix element of $\hat{O}_{{\rm Schiff}}^{{\rm atomic}}$
appearing in Eq.\,(\ref{eq:Schiff-E1a1}):
\begin{align}
\braket{n|\hat{O}_{\mathrm{Schiff}}^{\mathrm{atomic}}|\mathrm{g.s.}}= & _{\mathcal{N}}\bra{\widetilde{\mathrm{g.s.}}}\otimes\braket{P,\lambda|\hat{O}_{\mathrm{Schiff}}^{\mathrm{atomic}}|S}\otimes\ket{\widetilde{\mathrm{g.s.}}}_{\mathcal{N}}\,,
\end{align}
and note that we use the convention of Eq.\,(\ref{eq:pol-wf}). Focusing on the first term in
$\hat{O}_{{\rm Schiff}}^{{\rm atomic}}$ that contains $\mathcal{C}_{0>}(x)$
we obtain an electronic matrix element
\begin{align}
\braket{P,\lambda|\hat{O}_{\mathrm{Schiff}}^{\mathrm{atomic}}|S}_{\textrm{1st term}}= & -\frac{4\,\pi\,\alpha}{Z}\,\sqrt{\frac{1}{3}}\,\int\, d^{3}x\,\psi_{P}^{*}(\bm x)\,\frac{1}{x^{2}}\,\psi_{S}(\bm x)\,\sum_{m}\, Y_{1}^{m}(\hat{x})\,\bm d_{\mathcal{N}}^{m*}\,\hat{\mathcal{C}}_{0>}(x)\nonumber \\
= & -\frac{4\,\pi\,\alpha}{\sqrt{3}\, Z}\,\sum_{m}\,\int_{0}^{\infty}\, dx\, u_{P}(x)^{*}\, u_{S}(x)\,\int\, d\Omega_{x}\, Y_{1}^{\lambda*}(\hat{x})\, Y_{1}^{m}(\hat{x})\, Y_{0}(\hat{x})\nonumber \\
& \times\bm d_{\mathcal{N}}^{m*}\,\int\, d^{3}y\,\theta(y-x)\,\hat{\rho}(\bm y)\, Y_{0}(\hat{y})\,.
\end{align}
The angular $d\Omega_{x}$ integral gives $\delta_{m\lambda}/\sqrt{4\pi}$.
To evaluate the radial integral we first write
\begin{align}
u_{P}(x)^{*}\, u_{S}(x) & =\sum_{k\ge1}\, c_{k}\, x^{k}\,,
\end{align}
where the $c_{k}$ are given by the appropriate products of the $a_{k}$
and $b_{k}$ (our procedure here is similar to Ref.\,\cite{Flambaum:2001gq}).
Interchanging the order of integration and evaluating the integral
over the electronic radial coordinate gives
\begin{align}
\braket{P,\lambda|\hat{O}_{\mathrm{Schiff}}^{\mathrm{atomic}}|S}_{\textrm{1st term}} & =-\frac{4\,\pi\,\alpha}{\sqrt{3}\, Z}\,\frac{1}{4\,\pi}\,\bm d_{\mathcal{N}}^{\lambda*}\,\int\, d^{3}y\,\hat{\rho}(\bm y)\,\sum_{k\ge1}\,\frac{c_{k}\, y^{k+1}}{k+1}\,.
\end{align}
Evaluating the second and third terms in the matrix element of $\hat{O}_{{\rm Schiff}}^{{\rm atomic}}$ in a similar way leads to
\begin{align}
\braket{P,\lambda|\hat{O}_{\mathrm{Schiff}}^{\mathrm{atomic}}|S}_{\textrm{terms 1--3}}=
& (4\,\pi\,\alpha)\,\left(\frac{\sqrt{3}}{4\,\pi}\right)\,\int\, d^{3}y\,\hat{\rho}(\bm y)\,\sum_{k\ge1}\,\frac{c_{k}\, y^{k+1}}{(k+1)(k+4)}\,\Bigg\{\bm y^{\lambda*}\nonumber \\
& -\frac{(k+4)}{3\, Z}\,\left(\bm d_{\mathcal{N}}^{\lambda*}-\frac{2\,(k+1)\,\sqrt{2\,\pi}}{(k+4)}\,\left[\bm d_{\mathcal{N}}\otimes Y_{2}(\hat{y})\right]_{1}^{\lambda*}\right)\Bigg\}\,.
\label{eq:schiff1}
\end{align}
Note that in this expression $\bm d_{\mathcal{N}}^{\lambda}$ is the
nuclear operator defined by Eq.\,(\ref{eq:dndef}).

The ``local'' nuclear Schiff moment operator $\hat{\bm S}_{L}^{\lambda}$ can be defined by requiring that
\begin{align}
-(4\,\pi\,\alpha)\,\braket{P,\lambda|\hat{\bm S}_{L}\cdot\bm\nabla\,\delta^{(3)}(\bm x)|S} 
& \equiv\braket{P,\lambda|\hat{O}_{\mathrm{Schiff}}^{\mathrm{atomic}}|S}_{\textrm{terms 1--3}}\,.
\label{eq:schiff2}
\end{align}
Evaluating the left side of Eq.\,(\ref{eq:schiff2}) yields
\begin{align}
-(4\,\pi\,\alpha)\,\braket{P,\lambda|\hat{\bm S}_{L}\cdot\bm\nabla\,\delta^{(3)}(\bm x)|S}
& =(4\,\pi\,\alpha)\,\left(\frac{\sqrt{3}}{4\,\pi}\right)\, c_{1}\,\hat{\bm S}_{L}^{\lambda*}\,.
\end{align}
Including for the moment only the first three terms in $\hat{O}_{{\rm Schiff}}^{{\rm atomic}}$
as in Eq.\,(\ref{eq:schiff1}) gives
\begin{align}
\hat{\bm S}_{L}^{\lambda}= & \int\, d^{3}y\,\hat{\rho}(\bm y)\,\sum_{k\ge1}\,\frac{c_{k}}{c_{1}}\,\frac{y^{k+1}}{(k+1)(k+4)}\,\Bigg\{\bm y^{\lambda}\nonumber \\
& -\frac{(k+4)}{3\, Z}\,\left(\bm d_{\mathcal{N}}^{\lambda}-\frac{2\,(k+1)\,\sqrt{2\,\pi}}{(k+4)}\,\left[\bm d_{\mathcal{N}}\otimes Y_{2}(\hat{y})\right]_{1}^{\lambda}\right)\Bigg\}\,.
\label{eq:schiff3}
\end{align}
Retaining only the leading-order term, \textit{i.e.}, $k=1$, on the
RHS of Eq.\,(\ref{eq:schiff3}) yields our result for what is conventionally
referred to as the Schiff moment operator, $\hat{\bm S}^{\lambda}$:
\begin{align}
\hat{\bm S}^{\lambda}= & \frac{1}{10}\,\int\, d^{3}y\,\hat{\rho}(\bm y)\, y^{2}\,\Bigg\{\bm y^{\lambda}-\frac{5}{3\, Z}\,\left(\bm d_{\mathcal{N}}^{\lambda}-\frac{4\,\sqrt{2\,\pi}}{5}\,\left[\bm d_{\mathcal{N}}\otimes Y_{2}(\hat{y})\right]_{1}^{\lambda}\right)\Bigg\}\,
.\label{eq:schiff4}
\end{align}
Note that nuclear matrix elements of $\hat{\bm S}_{L}^{\lambda}$
and $\hat{\bm S}^{\lambda}$ will involve nuclear matrix elements
of of density-density correlations, since the last two terms on the
RHS of Eqs.\,(\ref{eq:schiff3}) and (\ref{eq:schiff4}) contain
the operator
\begin{align}
\bm d_{\mathcal{N}}^{\lambda}= & \int\, d^{3}z\,\hat{\rho}(\bm z)\,\bm z^{\lambda}\,,
\end{align}
that multiplies the operators $\hat{\rho}(\bm y)\, y^{2}$ or $\hat{\rho}(\bm y)\, y^{2}\, Y_{2}^{M}(\hat{y})$.

The form of the operators in Eqs.\,(\ref{eq:schiff3}) and (\ref{eq:schiff4})
differ in two important respects from the operators previously used
in the literature: (a) the presence of the final term involving the
$Y_{2}(\hat{y})$ and (b) the present treatment of $\bm d_{\mathcal{N}}^{\lambda}$
as an operator rather than as a $c$-number. To illustrate the importance
of this first difference, we consider the Schiff moment operator of the deuteron in the limit
where only the nuclear polarization term (due to $\tilde{V}^{\mathrm{nuc}}_{\mathrm{int}}$)
is retained.  That is, we ignore the effects of PVTV one-body ({\it e.g.}, the neutron and proton EDMs) and two-body charges and three-currents.  Polarization terms (pol.) contribute via the one-body
PCTC nuclear charge operator:
\begin{align}
\hat{\rho}(\bm y)=\sum_{k}\,\delta^{(3)}(\bm y-\bm r_{k})\,\tau_{p}^{k}\,,\qquad & \tau_{p}^{k}=\frac{1}{2}\,(1+\tau_{3}^{k})\,,
\label{eq:rho1}
\end{align}
where the sum is over both nucleons ($k=1,2$) with coordinate $r_{k}$.
Since only the proton contributes, we label the
proton coordinate $r$ for simplicity. Substituting Eq.\,(\ref{eq:rho1}) into Eq.\,(\ref{eq:schiff4}) leads to
\begin{align}
\hat{\bm S}^{\lambda}\vert_{\mathrm{deuteron}}^{\mathrm{pol.}}= & \frac{1}{10}\,\tau_{p}\, r^{2}\,\left\{ \left(1-\frac{5}{3}\right)\,\bm r^{\lambda}+\frac{4\,\sqrt{2\,\pi}}{3}\,\left[\bm r\otimes Y_{2}(\hat{r})\right]_{1}^{\lambda}\right\} \,,
\label{eq:schiff5}
\end{align}
where we have performed the integrals over $\bm y$ and $\bm z$.
Using
\begin{align}
\left[\bm r\otimes Y_{2}(\hat{r})\right]_{1}^{\lambda}= & -\frac{1}{\sqrt{2\,\pi}}\,\bm r^{\lambda}\,,
\end{align}
we obtain
\begin{align}
\hat{\bm S}^{\lambda}\vert_{\mathrm{deuteron}}^{\mathrm{pol.}}= & -\frac{1}{5}\,\tau_{p}\, r^{2}\,\bm r^{\lambda}\,,
\end{align}
where the coefficient $-1/5$ is a factor of three larger than had
we omitted the $Y_{2}(\hat{y})$ term. Generalizing this argument
to an arbitrary nucleus we obtain
\begin{align}
\hat{\bm S}^{\lambda}\vert^{\mathrm{pol.}}= & \frac{1}{10}\,\sum_{k}\,\tau_{p}^{k}\, r_{k}^{2}\,\left\{ \bm r_{k}^{\lambda}-\frac{5}{3\, Z}\,\sum_{j}\,\tau_{p}^{j}\,\left(\bm r_{j}^{\lambda}-\frac{4\,\sqrt{2\,\pi}}{5}\,\left[\bm r_{j}\otimes Y_{2}(\hat{r}_{k})\right]_{1}^{\lambda}\right)\right\} \,.
\label{eq:schiff6}
\end{align}
Focusing on the terms in Eq.\,(\ref{eq:schiff6}) for which $j=k$
and using $\tau_{p}^{k}\,\tau_{p}^{k}=\tau_{p}^{k}$ gives
\begin{align}
\hat{\bm S}^{\lambda}\vert_{j=k}^{\mathrm{pol.}}= & \frac{1}{10}\,\sum_{k}\,\tau_{p}^{k}\, r_{k}^{2}\,\bm r_{k}^{\lambda}\left\{ 1-\frac{5}{3\, Z}\,\left(1+\frac{4}{5}\right)\right\} \,.
\label{eq:schiff7}
\end{align}
where the final $4/5$ arises from the $Y_{2}(\hat{y})$ term. For
heavy nuclei, the effect of this correction will be suppressed by
the factor of $1/Z$ but its importance relative to the second term
in Eq.\,(\ref{eq:schiff4}) is of the same order. The ground-state expectation
value of this odd-parity operator will be nonzero because of PVTV admixtures 
in the nuclear wave function.

The presence of the terms with $j\not=k$ point to the impact of treating
$\bm d_{\mathcal{N}}$ as an operator rather than as a $c$-number.
To see this, we write $\hat{\bm S}^{\lambda}$ as follows:
\begin{align}
\hat{\bm S}^{\lambda}= & \frac{1}{10}\,\int\, d^{3}y\,\int\, d^{3}z\,\hat{\rho}(\bm y)\, y^{2}\,\Bigg\{\bm y^{\lambda}\,\delta^{(3)}(\bm z)-\frac{5}{3\, Z}\,\hat{\rho}(\bm z)\,\left(\bm z^{\lambda}-\frac{4\,\sqrt{2\,\pi}}{5}\,\left[\bm z\otimes Y_{2}(\hat{y})\right]_{1}^{\lambda}\right)\Bigg\}\,.
\label{eq:schiff8}
\end{align}
The product of charge density operators appearing in Eq.\,(\ref{eq:schiff8})
implies that nuclear matrix elements of $\hat{\bm S}^{\lambda}$
involve a two-body correlation. A comparison of this result with one
where $\bm d_{\mathcal{N}}$ is treated as a $c$-number
can be made by inserting a complete set of intermediate nuclear states
\begin{align}
_{\mathcal{N}}\braket{\widetilde{\mathrm{g.s.}}|\hat{\rho}(\bm y)\,\hat{\rho}(\bm z)|\widetilde{\mathrm{g.s.}}}_{\mathcal{N}}= & \sum_{n}\,{}_{\mathcal{N}}\braket{\widetilde{\mathrm{g.s.}}|\hat{\rho}(\bm y)|n}\braket{n|\hat{\rho}(\bm z)|\widetilde{\mathrm{g.s.}}}_{\mathcal{N}}\,.
\end{align}
Treating $\bm d_{\mathcal{N}}$ as a $c$-number amounts
to the assumption that this sum is effectively saturated by retaining
only $\ket{n}=\ket{\mathrm{g.s.}}$. The numerical validity of this
assumption is not at all clear, and certainly should be explored in
numerical calculations. This task goes beyond the scope of the present
paper, but will be explored in future work.

Additional contributions to the nuclear Schiff moment arise from S-P
matrix elements of the last four terms in $\hat{O}_{{\rm Schiff}}^{{\rm atomic}}$.
It is useful to distinguish contributions from these terms that are
sensitive to electronic penetration inside the nucleus from those
that are not. Following similar arguments to those used in deriving
$\hat{\bm S}_{L}$ we obtain
\begin{align}
\braket{P,\lambda|\hat{O}_{\mathrm{Schiff}}^{\mathrm{atomic}}|S}_{\textrm{terms 4--7}}\equiv & -(4\,\pi\,\alpha)\,\bra{P,\lambda}\bm\Delta\hat{\bm S}_{(1)}\cdot\delta^{(3)}(\bm x)\,\overleftrightarrow{\bm\nabla}+\bm\Delta\hat{\bm S}_{(2)}\dot{\odot}(\frac{1}{x^{3}}\, Y_{2}(\hat{x})\,\overleftrightarrow{\bm\nabla})\ket{S}\nonumber \\
& +\bra{P,\lambda}\bm\Delta\hat{\bm S}_{(3)}\cdot\overleftrightarrow{\bm\nabla}\ket{S}\,,
\label{eq:Oschiff4-7}
\end{align}
where
\begin{align}
\bm\Delta\hat{\bm S}_{(1)}^{\lambda}= & -\frac{1}{3\, Z}\,\sum_{j\ge1,k\ge0}\,\left(\frac{a_{k}\, b_{j}}{a_{0}\, b_{1}}\right)\,\int\, d^{3}z\,\int\, d^{3}y\, y^{j+k+1}\,\left[\hat{\rho}(\bm z)\,,\,\hat{\rho}(\bm y)\right]\nonumber \\
& \times\left\{ \frac{(j+2-k)}{(j+k+2)(j+k+1)}\,\bm z^{\lambda}-\frac{2\,\sqrt{2\,\pi}\,(j-1-k)}{(j+k+4)(j+k-1)}\,\left[Y_{2}(\hat{y})\otimes\bm z\right]_{1}^{\lambda}\right\} \,,\label{eq:schiff9a}\\
\bm\Delta\hat{\bm S}_{(2)}^{\lambda}= & -\frac{1}{5\, Z}\,\left[\bm d_{\mathcal{N}}^{\lambda}\,,\,\hat{C}_{2}\right]\,,\label{eq:schiff9b}\\
\bm\Delta\hat{\bm S}_{(3)}^{\lambda}= & \frac{1}{Z}\,\left[\bm d_{\mathcal{N}}^{\lambda}\,,\, H^{\mathrm{nuc}}_{\mathrm{int}}\right]\,.
\label{eq:schiff9c}
\end{align}
Details pertaining to the derivation of $\bm\Delta\hat{\bm S}_{(1)}^{\lambda}$
are given in the Appendix\,\ref{sec:deltas}; as for $\bm\Delta\hat{\bm S}_{(2,3)}^{\lambda}$,
they can be readily read by comparing Eqs.\,(\ref{eq:Schiff-E1a1})
and (\ref{eq:Oschiff4-7}).

The operator $\bm\Delta\hat{\bm S}_{(1)}^{\lambda}$ characterizes
corrections to Schiff screening that arise when the nuclear EDM operator
does not commute with local (penetration), even-$J$ moments of the
charge operator. The result is the presence of the commutator $[\hat{\rho}(\bm z)\,,\,\hat{\rho}(\bm y)]$
in Eq.\,(\ref{eq:schiff9a}). We expect that contributions from this
commutator will be suppressed by powers of $v/c$, where $v$ is a
typical nucleon velocity, since the leading-order (in $v/c$) parts
of the one-body charge operator give a vanishing commutator. Inclusion
of sub-leading one-body or two-body charge operators will not, in
general, give a vanishing commutator since these sub-leading terms
contain momentum and/or Pauli spin operators. Again, we defer a detailed
analysis of the numerical importance of these effects to a future
study and concentrate here on delineating the various contributions
to the operator.

The operators $\bm\Delta\hat{\bm S}_{(2,3)}^{\lambda}$ characterize
long-distance screening corrections that occur when the nuclear EDM
operator does not commute with either the quadrupole moment operator
or the internal nuclear Hamiltonian. We expect the former to be suppressed
by powers of $v/c$ for similar reasons as for $\bm\Delta\hat{\bm S}_{(1)}^{\lambda}$.
The commutator of $\bm d_{{\cal N}}^{\lambda}$ with $H^{{\rm nuc}}_{{\rm int}}$
is, perhaps, more subtle. In general, the nuclear Hamiltonian contains
both momentum- and spin-dependent forces. The commutator of the momentum
and spin operators will not commute with $\bm d_{{\cal N}}^{\lambda}$
which generally contains both spin- and space-dependent components.
As with $\bm\Delta\hat{\bm S}_{(1)}^{\lambda}$ we defer a detailed
analysis of these corrections to a future study.

\subsection*{Magnetic corrections}

Additional corrections to Schiff screening arise from the magnetic
operators in $\hat{O}_{3,4,8,9}^{(\mathcal{N})}$. As with the Schiff
operator terms, we collect the lowest rank terms which can also induce electronic S-P transitions:
\begin{eqnarray}
\left[\hat{O}_{3,4,8,9}^{(\mathcal{N})}\right]_{{\rm lowest\  rank}} & = & -\frac{4\,\pi\,\alpha}{Z\, x^{3}}\,\Bigg\{[Y_{2}(\hat{x})\otimes\bm\alpha]_{1}\odot\left[\bm d_{{\cal N}}\otimes\sqrt{\frac{1}{6}}\,\left(\hat{M}_{1}-\hat{\mathcal{M}}_{1>}(x)\right)\right]_{1}\nonumber \\
&  & -[Y_{2}(\hat{x})\otimes\bm\alpha]_{2}\odot\left[\bm d_{{\cal N}}\otimes\left(\sqrt{\frac{3}{10}}\,\left(\hat{M}_{1}+\hat{\mathcal{M}}_{1>}(x)\right)+\sqrt{\frac{8}{15}}\,\hat{\mathcal{M}}_{1<}(x)\right)\right]_{2}\nonumber \\
&  & -[Y_{0}(\hat{x})\otimes\bm\alpha]_{1}\odot\left[\bm d_{{\cal N}}\otimes\sqrt{\frac{1}{3}}\,\hat{\mathcal{M}}_{1<}(x)\right]_{1}\nonumber \\
&  & +[Y_{2}(\hat{x})\otimes\bm\alpha]_{2}\odot\frac{1}{5}\,\left(\hat{M}_{2}+\hat{\mathcal{M}}_{2}(x)\right)\nonumber \\
&  &  +(x\,[Y_{1}(\hat{x})\otimes\bm\alpha]_{1}\,\overleftrightarrow{\bm\nabla})\dot{\odot}\left[\bm d_{{\cal N}}\,,\,\frac{1}{3}\,\left(\hat{M}_{1}+\hat{\mathcal{M}}_{1}(x)\right)\right]\Bigg\}\,\equiv\,\hat{O}_{{\rm Schiff-mag}}^{{\rm atomic}}\,.
\label{eq:Schiff-M1a}
\end{eqnarray}
The operator $\hat{O}_{{\rm Schiff-mag}}^{{\rm atomic}}$ contains
two types of terms: (a) those arising from the ``displacement''
of the magnetic dipole interaction, containing the commutator of $\bm d_{\mathcal{N}}$
with the magnetic dipole operators, and (b) an explicit magnetic quadrupole
interaction. Effects of the former type were identified in Schiff's
original paper, but have generally been neglected in subsequent work.
For atoms with nuclear spins $I=1/2$, there is no magnetic quadrupole contribution.

The form of $\hat{O}_{{\rm Schiff-mag}}^{{\rm atomic}}$ simplifies
when we consider the electronic matrix element, again under the factorization
assumption of Eq.\,(\ref{eq:fact}). To arrive at this simplification,
we consider the symmetry properties of the relevant electronic operator
\begin{align}
\mathcal{O}_{j}^{m}\equiv & \left[Y_{l}(\hat{x})\otimes\bm\alpha\right]_{j}^{m}\,.
\label{eq:mag1}
\end{align}
To proceed further, it is useful to consider the perturbed atomic
states appearing in Eq.\,(\ref{eq:deltaE2screen}) and label them
according to their angular momentum quantum numbers. We define the
states $\ket{\tilde{J}^{\prime}M_{J}=J}_{e}$ entering Eq.\,(\ref{eq:deltaE2screen})
via
\begin{align}
\sum_{J^{'}}\,\ket{\tilde{J}^{'}J}_{e}\otimes\ket{\mathrm{g.s}}_{\mathcal{N}}\equiv & \sum_{n}\,\frac{1}{E_{\mathrm{g.s.}}-E_{n}}\,_e\braket{n|V_{\mathrm{ext}}^{(e)}|JJ}_e\,
\ket{n}_e \otimes \ket{\mathrm{g.s}}_{\mathcal{N}}\,,
\label{eq:mag3}
\end{align}
where $\ket{J,J}_{e}$ denotes the electronic ground state and $V_{{\rm ext}}^{(e)}$
contains the potential associated with constant external electric
field that mixes electronic states differing in orbital angular momentum
by one unit. The electronic component of Eq.\,(\ref{eq:deltaE2screen})
involving a generic $\mathcal{O}_{j}^{m}$ operator then has the form
\begin{align}
\left[_{e}\braket{\tilde{J}^{'}J|\mathcal{O}_{j}^{0}|JJ}_{e}+_{e}\braket{JJ|\mathcal{O}_{j}^{0}|\tilde{J}^{'}J}_{e}\right]\,.
\label{eq:mag4}
\end{align}

Now using the following identity
\begin{equation}
\braket{a|\mathcal{O}|b}=\braket{T(b)|T\,\mathcal{O}^{\dagger}\, T^{-1}|T(a)}\,,
\end{equation}
where $T$ is the anti-unitary time-reversal operator, along with the Hermiticity and time-reversal properties of $\mathcal{O}_{j}^{m}$
that
\begin{align}
[Y_{l}\otimes\bm\alpha]_{j}^{m\dagger}=(-1)^{l+1-j+m}\,[Y_{l}\otimes\bm\alpha]_{j}^{-m}\,, & \qquad T\,[Y_{l}\otimes\bm\alpha]_{j}^{m}\, T^{-1}=(-1)^{l-j+m}\,[Y_{l}\otimes\bm\alpha]_{j}^{m}\,,
\end{align}
we can show that
\begin{align}
_{e}\braket{\tilde{J}^{'}J|\mathcal{O}_{j}^{0}|JJ}_{e} & =-{}_{e}\braket{T(JJ)|\mathcal{O}_{j}^{0}|T(\tilde{J}^{'}J)}_{e}\nonumber \\
& =-(-1)^{J+J+L}\,(-1)^{J^{'}+J+L^{'}}\braket{J-J|\mathcal{O}_{j}^{0}|\tilde{J}^{'}-J}\nonumber \\
& =-(-1)^{J^{'}-J+L^{'}+L}\,(-1)^{2\, J+J^{'}+J+j}\,\braket{JJ|\mathcal{O}_{j}^{0}|\tilde{J}^{'}J}\,,
\end{align}
where the second line follows from the phase convention for the time-reversed
state (we use here the Condon-Shortely convention\,%
\footnote{Another popular phase convention is the one of Biedenharn-Rose, which
introduces an extra {}``$i^{L}$'' factor for $Y_{L}^{M}$. But
the result is independent of phase convention.} %
), and the third line follows from the symmetry properties of the $3-j$
symbol. Given the fact that $(L^{'}+L)$ is odd (parity change) and
$2\,(J^{'}+J)$ is even, we have
\begin{align}
_{e}\braket{\tilde{J}^{'}J|\mathcal{O}_{j}^{0}|JJ}_{e} & =(-1)^{j}\,\braket{JJ|\mathcal{O}_{j}^{0}|\tilde{J}^{'}J}\,,
\end{align}
which means the sum, Eq.\,(\ref{eq:mag4}), can only be nonvanishing
for $j=\mathrm{even}$. Thus, the $[Y_{0,2}(\hat{x})\otimes\bm\alpha]_{1}$
parts in Eq.\,(\ref{eq:Schiff-M1a}) yield no contribution to the
induced atomic dipole moment and we may work with the operator
\begin{eqnarray}
\hat{O}_{{\rm Schiff-mag}}^{{\rm atomic}} & = & -\frac{4\,\pi\,\alpha}{Z\, x^{3}}\,\Bigg\{\nonumber \\
&  & -[Y_{2}(\hat{x})\otimes\bm\alpha]_{2}\odot\left[\bm d_{{\cal N}}\otimes\left(\sqrt{\frac{3}{10}}\,\left(\hat{M}_{1}+\hat{\mathcal{M}}_{1>}(x)\right)+\sqrt{\frac{8}{15}}\,\hat{\mathcal{M}}_{1<}(x)\right)\right]_{2}\nonumber \\
&  & +[Y_{2}(\hat{x})\otimes\bm\alpha]_{2}\odot\frac{1}{5}\,\left[\hat{M}_{2}+\hat{\mathcal{M}}_{2}(x)\right]\nonumber \\
&  & +(x\,[Y_{1}(\hat{x})\otimes\bm\alpha]_{1}\,\overleftrightarrow{\bm\nabla})\dot{\odot}\left[\bm d_{{\cal N}}\,,\,\frac{1}{3}\,\left(\hat{M}_{1}+\hat{\mathcal{M}}_{1}(x)\right)\right]\Bigg\}\,.
\label{eq:Schiff-M1c}
\end{eqnarray}

Several comments can be made regarding this $\hat{O}_{{\rm Schiff-mag}}^{{\rm atomic}}$
term:
\begin{enumerate}
\item As $\hat{O}_{{\rm Schiff-mag}}^{{\rm atomic}}$ involves interactions
with static magnetic moments $M_{1}$ (through the combination with
$\bm d_{\mathcal{N}}$) and $M_{2}$, they are not affected by the
screening. Compared to the finite-size effects which contain a suppression
factor roughly to the order of $\textrm{(nuclear size)}^{2}/\textrm{(atomic size)}^{2}\sim\mbox{fm}^{2}/a_{0}^{2}\sim10^{-9}$,
these magnetic effects are only suppressed by the typical hyperfine
factor $\alpha^{2}\, m_{e}/m_{N}\sim10^{-7}$ (note that we have not
taken any atomic or nuclear enhancement factor into account, just
use a pure dimensional analysis). These could be potentially important,
as already pointed out by Schiff\,\cite{Schiff:1963} and others
(see, \textit{e.g.},\,\cite{Khriplovich:1997,Sandars:2001nq}), particularly
for the open-shell atoms.
\item The last term in $\hat{O}_{{\rm Schiff-mag}}^{{\rm atomic}}$, unlike
the rest which are of quadrupole nature and need a nuclear spin equal
or greater than $1$, contributes for any nucleus with spin. In fact,
it has been identified in the original paper by Schiff\,\cite{Schiff:1963}
(he considered the hydrogen atom only) and later on studied by Hinds
and Sandars\,\cite{Hinds:1980a} with a more refined expression.
The latter authors find a non-negligible contribution from this term,
about $20\%$, to the EDM of $\mbox{TlF}$ molecule, assuming the
proton EDM is the only CPV source. We emphasize that this term is
a result of taking $\bm d_{\mathcal{N}}$ and the magnetic dipole operators ${\hat M}_1$, $\hat{\mathcal{M}}_1$
as \emph{operators}: had one started with $\bm d_{\mathcal{N}}$ as
a $c$-number, this term would not have existed at all.
\end{enumerate}

\section{Practical Considerations\,\label{sec:practical}}

In the foregoing analysis, we have laid out the structure of the atomic
operators that describe the corrections to Schiff screening and that
characterize the leading contributions to an atomic EDM arising from
the relativistic motion of the electrons, the finite size and internal
structure of the nucleus, and magnetic interactions between the atomic
electrons and nucleus. In doing so, we have attempted to remain as
general as possible without making specific reference to the atomic
states. In this context, the leading contribution associated with
finite nuclear size and internal structure is given by $\hat{O}_{{\rm Schiff}}^{{\rm atomic}}$
in Eq.\,(\ref{eq:Schiff-E1a1}). Given a complete basis of atomic
states including the effects of P- and T-odd admixtures into the nuclear
states of definite parity -- one could in principle use $\hat{O}_{{\rm Schiff}}^{{\rm atomic}}$
to compute the energy shift of an atom in an external electric field
\textit{via} Eq.\,(\ref{eq:deltaE2screen}).

As a practical matter, it has been the convention to specify the computation
of finite-size corrections by considering the effect of $\hat{O}_{{\rm Schiff}}^{{\rm atomic}}$
on mixing between electronic S- and P- electronic states, making the
factorization assumption of Eq.\,(\ref{eq:fact}) and neglecting
nuclear polarization corrections. In order to compare the implications
of our formulation with previous analyses, we have derived an effective
nuclear Schiff moment operator from $\hat{O}_{{\rm Schiff}}^{{\rm atomic}}$
under these assumptions and break down the matrix element as
\begin{align}
\braket{P,\lambda|\hat{O}_{{\rm Schiff}}^{{\rm atomic}}|S}= & (-4\,\pi\,\alpha)\bra{P,\lambda}\Big\{\hat{\bm S}_{L}\cdot\bm\nabla\,\delta^{(3)}(\bm x)+\bm\Delta\hat{\bm S}_{(1)}^{\lambda}\cdot\delta^{(3)}(\bm x)\,\overleftrightarrow{\bm\nabla}\nonumber \\
& +\bm\Delta\hat{\bm S}_{(2)}^{\lambda}\dot{\odot}\frac{1}{x^{3}}\, Y_{2}(\hat{x})\,\,\overleftrightarrow{\bm\nabla}\Big\}\ket{S}+\bra{P,\lambda}\bm\Delta\hat{\bm S}_{(3)}\cdot\overleftrightarrow{\bm\nabla}\ket{S}\,,
\end{align}
where $\hat{\bm S}_{L}^{\lambda}$ and $\bm\Delta\hat{\bm S}_{(1-3)}^{\lambda}$
are given by Eqs.\,(\ref{eq:schiff3}) and (\ref{eq:schiff9a}--\ref{eq:schiff9c})
respectively. We emphasize that different effective operators will
apply for other atomic transitions. The derivation of the relevant
operators starting from $\hat{O}_{{\rm Schiff}}^{{\rm atomic}}$ will
follow the similar logic as in our derivation here for $\braket{P,\lambda|\hat{O}_{{\rm Schiff}}^{{\rm atomic}}|S}$.
One can introduce a further level of approximation by specifying the
sums in these equations to the leading terms, corresponding to retaining
only the leading $x$-dependence of the electronic wave function near
the origin. Quantifying the error introduce by making either the factorization
approximation or retaining only the leading electronic $x$-dependence
is an important task for future nuclear and atomic structure computations.

Having introduced these two approximations, we have obtained a form
for $\braket{P,\lambda|\hat{O}_{{\rm Schiff}}^{{\rm atomic}}|S}$
that differs in several respects from the nuclear Schiff moment operator
used previously in the literature. These differences arise primarily
because we have reformulated Schiff's theorem entirely at the operator
level, have retained the full nuclear operator-dependence of the finite-size
correction, and proceeded consistently within the framework of
the spherical multipole expansion. In brief, the resulting differences
with previous forms of the operator obtained in our approach are:

\begin{itemize}
\item [(i)] The presence of the $Y_{2}(\hat{x})$ term and the product
of nuclear density \textit{operators} in Eqs.\,(\ref{eq:schiff3}, \ref{eq:schiff4}); 
\item [(ii)] The presence of the commutator of nuclear density operators
in Eqs.\,(\ref{eq:schiff9a}, \ref{eq:schiff9b}) that vanishes at
leading order in $v/c$ but will not in general vanish at higher orders
due to the presence of spin- and momentum-dependences in the nuclear
charge operators;
\item [(iii)] The presence of the internal nuclear Hamiltonian in Eq.\,(\ref{eq:schiff9c})
that results from including the internal nuclear degrees of freedom
as dynamical quantities in the atomic Hamiltonian.
\end{itemize}
We have illustrated the potential impact of including some of these
new ingredients by studying the effect of the $Y_{2}(\hat{x})$ term
in $\hat{\bm S}^{\lambda}$ in the deuteron, where it increases the
magnitude of the Schiff moment by a factor of three at the one-body
operator level, and in the diagonal part of the density-density operator
matrix element for heavy nuclei, where its impact is $1/Z$ suppressed.
We have not, however, quantified the effect of non-diagonal terms
and the corresponding nuclear excitations, those that arise at higher-order
in $v/c$, or those associated with the internal nuclear Hamiltonian.
The result of future studies that quantify these contributions will
determine the degree to which previous computations of the Schiff
moment adequately characterize the leading nuclear correction to Schiff
screening.

Additional corrections arise from the effect of magnetic interactions
between the atomic electrons and nucleus. In principle, these corrections
could be important for the theoretical interpretation of other atomic
EDMs. To that end, we have worked out the leading form of the magnetic
operator $\hat{O}_{{\rm Schiff-mag}}^{{\rm atomic}}$, again in a
way that makes no reference to the atomic states and later specifying
to the situation of a simple direct product of nuclear and electronic
states. Analyzing the quantitative impact of this operator will also
be an interesting endeavor for future nuclear and atomic structure
studies.

\begin{acknowledgments}
We thank V. V. Flambaum and P. Vogel for useful discussions. Part of
this work was supported by (i) the Dutch Stichting voor
Fundamenteel Onderzoek der Materie (FOM) under program 48, TRI$\mu$P (CPL and RGET), (ii)
U.S. Department of Energy under Contracts DE-AC52-06NA25396 (CPL),
DE-FG02-05ER41361 (MJR-M), and DE-FG02-00ER-41132 (WCH), and (iii) National Science Foundation under grant PHY-0555674 (MJR-M).
\end{acknowledgments}
\appendix

\section{Symmetrization of $[A(\bm y)\, B(\bm x)\,,\, C(\bm y)\, D(\bm x)]$\,\label{sec:symmetrization}}

When evaluating Eq.\,(\ref{eq:dN-final}) for the residual internal
PVTV $e$-$\mathcal{N}$ interaction that evades the Schiff screening
for $d_{\mathcal{N}}$, the commutator $[\bm d_{\mathcal{N}}\cdot\bm\nabla\,,\,\Delta H_{0}]$,
can be generically expressed as $[A(\bm y)\, B(\bm x)\,,\, C(\bm y)\, D(\bm x)]$
with $\mathcal{O}(\bm y)$ and $\mathcal{O}(\bm x)$ denoting operators acting upon nuclear
and electronic Hilbert spaces, respectively.  In order to obtain a manifestly Hermitian result, we employ the following identity:
\begin{align}
[A(\bm y)\, B(\bm x)\,,\, C(\bm y)\, D(\bm x)]= & \nicefrac{1}{2}\,\Big([B(\bm x)\,,D(\bm x)]\,\{ A(\bm y)\,,\, C(\bm y)\}\nonumber \\
& +\{ B(\bm x)\,,D(\bm x)\}\,[A(\bm y)\,,\, C(\bm y)]\Big)\,,
\end{align}
where we have used the fact that operators depending on $\bm x$ and $\bm y$ commute: $[A(\bm y)\,,\,B(\bm x)\, ] = 0$, {\em etc}.
As an illustrative example, we go through the terms in $\Delta H_{0}$
which involve static charge multipoles $C_{J}$'s with $J\ge2$. For
this case, one takes

\begin{align}
A(\bm y)=\bm d_{\mathcal{N}}\,, & \qquad B(\bm x)=\bm\nabla\,,\nonumber \\
C(\bm y)=C_{J}^{M}\,, & \qquad D(\bm x)=\frac{1}{x^{J+1}}\, Y_{J}^{M*}(\hat{x})\,,
\end{align}
and uses the gradient formula
\begin{equation}
[\bm\nabla_{x}\,,\,\frac{1}{x^{J+1}}\, Y_{J}^{M*}(\hat{x})]=\sqrt{(J+1)\,(2\, J+1)}\,\frac{1}{x^{J+2}}\,\bm Y_{JJ+11}^{M*}(\hat{x})\,,
\end{equation}
where $\bm Y_{JL1}^{M}$ is the vector spherical harmonics, then the
commutator can be reduced as
\begin{align}
-\left[\bm d_{\mathcal{N}}\cdot\bm\nabla_{x}\,,\, C_{J}\odot\frac{1}{x^{J+1}}\, Y_{J}(\hat{x})\right]= & \sqrt{\frac{J+1}{2\, J+3}}\,\frac{(2\, J+1)}{x^{J+2}}\, Y_{J+1}(\hat{x})\odot[\bm d_{\mathcal{N}}\otimes C_{J}]_{J+1}^{\mathrm{(sym)}}\nonumber \\
& +\frac{1}{x^{J+1}}\, Y_{J}(\hat{x})\,\overleftrightarrow{\bm\nabla}\dot{\odot}[\bm d_{\mathcal{N}}\,,\, C_{J}]\,,
\end{align}
with the short-hand notations $[...]^{\mathrm{sym}}$, $\overleftrightarrow{\bm\nabla}$,
and $\dot{\odot}$, being defined in Eqs.\,(\ref{eq:sym_n}--\ref{eq:dbldot}). This
result leads to the $\hat{O}_{1}^{(\mathcal{N})}$ operator, \textit{i.e.},
Eq.\,(\ref{eq:O1}).

\section{Derivation of $\bm\Delta\hat{\bm S}_{(1)}$\,\label{sec:deltas}}

In this appendix, we use the term involving $\mathcal{C}_{0}(x)$
in Eq.\,(\ref{eq:Schiff-E1a1}), {\it i.e.}, the $4$th term, to illustrate
the general procedure for obtaining $\bm\Delta\hat{\bm S}_{(1)}$.
First, the electronic matrix element
\begin{align}
\braket{P,\lambda|\hat{O}_{\mathrm{Schiff}}^{\mathrm{atomic}}|S}_{\textrm{term 4}}= & \braket{P,\lambda|\frac{4\,\pi\,\alpha}{Z\, x}\, Y_{0}(\hat{x})\otimes[\bm d_{\mathcal{N}}\,,\mathcal{C}_{0}(x)]\cdot\overleftrightarrow{\bm\nabla}]|S}\,,
\end{align}
can be easily evaluated with
\begin{align}
& \int\, d^{3}x\,\frac{1}{x}\, Y_{0}(\hat{x})\,\theta(y-x)\,(\frac{x}{y}-1)\,\frac{1}{2}\,\left(\psi_{P,\lambda}^{*}\,\bm\nabla\psi_{S}-\psi_{S}\,\bm\nabla\psi_{P,\lambda}\right)\nonumber \\
& =\frac{\bm e_{\lambda}^{*}}{4\,\sqrt{3\,\pi}}\sum_{j\ge1,k\ge0}\frac{(j-k+2)}{(j+k+2)(j+k+1)}\, b_{j}\, a_{k}\, y^{j+k+1}\,,
\end{align}
so that
\begin{align}
& \braket{P,\lambda|\hat{O}_{\mathrm{Schiff}}^{\mathrm{atomic}}|S}_{\textrm{term 4}}\nonumber \\
& =\frac{\alpha}{2\,\sqrt{3}\, Z}\sum_{j\ge1,k\ge0}b_{j}\, a_{k}\,\int\, d^{3}z\,\int\, d^{3}y\, y^{j+k+1}\,[\hat{\rho}(\bm z)\,,\,\hat{\rho}(\bm y)]\,\frac{(j-k+2)}{(j+k+2)(j+k+1)}\,\bm z^{\lambda*}\,.
\end{align}
Since this specific matrix element has been defined as (see Eq.\,(\ref{eq:Oschiff4-7}))
\begin{align}
\braket{P,\lambda|\hat{O}_{\mathrm{Schiff}}^{\mathrm{atomic}}|S}_{\textrm{terms 4,6}} & \equiv-(4\,\pi\,\alpha)\,\braket{P,\lambda|\bm\Delta\hat{\bm S}_{(1)}\odot\delta^{(3)}(\bm x)\overleftrightarrow{\bm\nabla}|S}\\
& =-(4\,\pi\,\alpha)\,\left(\frac{\sqrt{3}}{4\,\pi}\right)\,\frac{1}{2}\, b_{1}\, a_{0}\,\hat{\bm S}_{(1)}^{\lambda*}\,,
\end{align}
then the contribution of the $\mathcal{C}_{0}$ term to $\hat{\bm S}_{(1)}^{\lambda}$
becomes
\begin{align}
\bm\Delta\hat{\bm S}_{(1),\textrm{term 4}}^{\lambda}= & -\frac{1}{3\, Z}\,\sum_{j\ge1,k\ge0}\left(\frac{b_{j}\, a_{k}}{b_{1}\, a_{0}}\right)\,\int\, d^{3}z\,\int\, d^{3}y\, y^{j+k+1}\,[\hat{\rho}(\bm z)\,,\,\hat{\rho}(\bm y)]\nonumber \\
& \times\,\frac{(j-k+2)}{(j+k+2)(j+k+1)}\,\bm z^{\lambda}\,.
\end{align}
Similar logic applies to the term involving $\mathcal{C}_{2}(x)$.

\bibliographystyle{apsrev}
\bibliography{Schiff.bbl}



\end{document}